\documentclass[aps,prl, twocolumn,superscriptaddress]{revtex4-2}

\usepackage{amsmath,amssymb}
\usepackage{amsthm}
\usepackage{graphicx}
\usepackage{xcolor}
\usepackage{here}
\usepackage{enumerate}
\usepackage{braket}
\usepackage{url}
\usepackage[normalem]{ulem}
\usepackage{mathtools}
\usepackage{algorithm}
\usepackage{algpseudocode}

\newcommand{\todai}{Department of Physics, Graduate School of Science, The University of Tokyo, Hongo 7-3-1, Bunkyo-ku, Tokyo 113-0033, Japan}
\newcommand{\nii}{National Institute of Informatics,
Hitotsubashi 2-1-2, Chiyoda-ku, Tokyo 101-8430, Japan}
\newcommand{\transscale} {Trans-Scale Quantum Science Institute, The University of Tokyo, Hongo 7-3-1, Bunkyo-ku, Tokyo 113-0033, Japan}
\newcommand{\perimeter}{Perimeter Institute for Theoretical Physics, 31 Caroline Street North, Waterloo, Ontario, N2L 2Y5, Canada}
\newcommand{\iqc}{Institute for Quantum Computing, University of Waterloo, 200 University Avenue West, Waterloo, Ontario, N2L 3G1, Canada}

\newtheorem{de}{Definition}
\newtheorem{lem}{Lemma}

\newtheorem{Theorem}{Theorem}
\newcommand{\Proof}{{\bf Proof. \,}}

\begin{document}
\title{Higher-order quantum transformations of Hamiltonian dynamics}
\author{Tatsuki Odake}
\affiliation{\todai}
\author{Hl\'er Kristj\'ansson}
\altaffiliation{Current address: \perimeter, and \iqc}
\affiliation{\todai}
\author{Akihito Soeda}
\affiliation{\nii}
\affiliation{\todai}
\author{Mio Murao}
\affiliation{\todai}
\affiliation{{\transscale}}
\begin{abstract}

We present a quantum algorithm to achieve higher-order transformations of Hamiltonian dynamics. Namely, the algorithm takes as input a finite number of queries to a black-box seed Hamiltonian dynamics to simulate a desired Hamiltonian.
Our algorithm efficiently simulates linear transformations of any seed Hamiltonian with a bounded energy range consisting of a polynomial number of terms in system size,
making use of only controlled-Pauli gates
and time-correlated randomness.
This algorithm is an instance of quantum functional programming, where the desired function is specified as a concatenation of higher-order quantum transformations.
By way of example, we demonstrate the simulation of negative time-evolution and time-reversal, and perform a Hamiltonian learning task.
\end{abstract}
\maketitle

Efficiently simulating the dynamics of complex quantum systems is often stated as one of the main motivations of quantum computing. While such simulation is considered hard on classical computers, a range of efficient quantum algorithms have been developed for simulating Hamiltonian dynamics \cite{suzuki1990fractal,suzuki1991general,campbell2019random,berry2015simulating,low2017optimal,low2019hamiltonian,childs2018toward}. 
The core principle behind the standard Hamiltonian simulation algorithms is that the desired Hamiltonian dynamics can be well-approximated by a series of (arguably) simpler quantum operations. These algorithms rely on having a classical description of the desired Hamiltonian, which can often be used for obtaining a decomposition into a sum of easily implementable terms. This limits the way we can develop large-scale, complex quantum programs for dynamics simulation. Quantum algorithms which do not require detailed descriptions of quantum resources have a higher flexibility in quantum software development. 
This is related  to the fundamental problem of understanding how much quantum algorithms need to rely on the classical description of their inputs in order to achieve quantum advantages in information processing.

In this work, we study Hamiltonian dynamics that can be implemented given a \textit{seed Hamiltonian} $H$ without using a classical description of $H$.
That is, we study transformations of black-box Hamiltonians.
We present a quantum algorithm that simulates the dynamics of $f(H)$, where $f$ is any physically realizable linear function of $H$, given a description of $f$ and using a 
black-box Hamiltonian $H$ with a bounded energy range. This algorithm is an instance of a \textit{higher-order quantum transformation} on the unitary operation realized by the seed Hamiltonian dynamics. The functions that the algorithm can implement include
both the negative time-evolution and the time-reversal of an unknown Hamiltonian evolution  
by considering $f(H)=-H$ and $f(H)=H^T$ (transposition of $H$ in terms of the computational basis), respectively.  
Such general transformations have applications ranging from fundamental physics simulations to potential improvements in state-of-the-art algorithms, such as the Hamiltonian singular value transformation \cite{lloyd2021hamiltonian}.  We also show an application of our algorithm for Hamiltonian learning \cite{granade2012robust}, in particular, a task of efficiently estimating a parameter of a multi-parameter Hamiltonian using Hamiltonian dynamics by appropriately choosing $f(H)$.

Our work constitutes the first systematic study of higher-order quantum transformations in the context of Hamiltonian dynamics. Higher-order quantum transformations have attracted significant attention in recent years in the context of quantum circuit transformations, and are also known as superchannels, supermaps, quantum combs and process matrices \cite{chiribella2008quantum,chiribella2008transforming,chiribella2009theoretical,bisio2019theoretical,chitambar2019quantum, oreshkov2012quantum}. 
Higher-order algorithms for quantum computation can be seen as an analogue of functional programming in classical computing, where the possible inputs to an algorithm are quantum channels (for example, unitaries) specified ``operationally'' by their input-output description only (i.e.\ as black boxes). 

Previous works on this topic have focused on the possible transformations that can be achieved when the input channels are taken to be a finite sequence of quantum gates \cite{chiribella2008quantum,miyazaki2019complex,dong2019controlled, quintino2019probabilistic,quintino2019reversing,yoshida2022reversing,chiribella2019quantum,chiribella2013quantum,pollock2018non,oreshkov2012quantum,bai2020efficient}. Yet, the resources available in a given computation are not always best described by a finite sequence of gates, but rather by a continuously parameterized Hamiltonian evolution. In fact, it is known that certain functions such as controllization, which cannot be implemented on black box unitaries  \cite{gavorova2020topological,araujo2014quantum,soeda2013limitations,friis2014implementing}, can in fact be implemented if access to the underlying Hamiltonian evolution is given \cite{nakayama2015quantum,dong2019controlled}. This is because it is possible to apply an arbitrary fractional power of an unknown Hamiltonian evolution by changing the evolution time, whereas applying a fractional power is not possible for black box unitaries. 

{\it Summary of algorithm.---}We now present our algorithm in detail (see Algorithm 1). We represent Hilbert spaces of an $n$-qubit quantum system and a single-qubit auxiliary system by $\mathcal{H}$ and $\mathcal{H}_c$, respectively.
We assume that we can invoke the Hamiltonian evolution $e^{-i H \tau}$ of a seed Hamiltonian $H \in \mathcal{L}(\mathcal{H})$ with an upper bound $\Delta_H$ of the difference between the maximum and the minimum energy eigenvalues is given, for any time $\tau>0$.

\begin{algorithm}[H]
    \caption{Simulating $e^{-if(H)t}$}
    \begin{algorithmic}[1]
        \label{alg:1}
        \Statex{\textbf{Input:}}
        \begin{itemize}
            \item A finite number of queries to a black box Hamiltonian dynamics $e^{-iH\tau}$ of a seed Hamiltonian $H$ with $\tau>0$  on an $n$-qubit system $\mathcal{H}$
            \item An upper bound $\Delta_{H}$ of the difference between the maximum and the minimum energy eigenvalues
            \item Hermitian-preserving linear map $f:\mathcal{L}(\mathcal{H})\to \mathcal{L}(\mathcal{H})$ satisfying  $f(I)\propto I$, which can always be represented by the Pauli transfer matrix elements $\gamma_{\vec{w},\vec{u}}$ as
            \vspace{-0.5em}
            \begin{equation}
            \label{ptm}
                f = \sum_{\substack{\vec{w} \in \{0,1,2,3\}^n \\ \vec{u} \in \{0,1,2,3\}^n\setminus (0,\ldots ,0)}}
                \hspace{-1em}
                \gamma_{\vec{w},\vec{u}}
                \,
                f_{\vec{w},\vec{u}} \, , 
                	\vspace{-0.5em}
            \end{equation}
        	for some $\gamma _{\vec{w},\vec{u}} \in \mathbb{R}$ and functions $f_{\vec{w},\vec{u}}$ defined by
        		\vspace{-0.5em}
       	\begin{equation}
        		f_{\vec{w},\vec{u}}
        		(\sigma_{\vec{v}}) :=
        		\delta_{\vec{v}, \vec{u}}
        		\sigma_{\vec{w}}  
        		\label{single_pauli_transfer}
       			\vspace{-0.5em}
        	\end{equation}
            for any tensor products of Pauli operators $\sigma_{\vec{v}}:=\sigma_{v_1}\otimes \cdots \otimes \sigma_{v_n}$, where $\sigma_0 \!=\!I,\ \sigma_1\!=\!X,\ \sigma_2\!=\!Y,\ \sigma_3\!=\!Z$ and $\vec{u},\ \vec{v},\ \vec{w} \in \{0,1,2,3\}^n$ are Pauli index vectors
            \item Input state $\ket{\psi}\in \mathcal{H}$,\ Allowed error $\epsilon >0$,\ Time $t>0$
        \end{itemize}
        \Statex{\textbf{Output:}}
        A state approximating $e^{-if(H)t}\ket{\psi}$ with an error less than $\epsilon$ (measured by the 1-norm)
    \Statex \hrulefill
    	\Statex{\textbf{Runtime:}}
    	 $O(\beta^2t^2\Delta_H^2n/\epsilon)$ for $\beta :=2\sum_{\vec{w}, \vec{u}}|\gamma_{\vec{w}, \vec{u}}|$
    	\Statex{\textbf{Used Resources:}}
    	\Statex \hskip1.0em System: $\mathcal{H}$ and one  auxiliary qubit $\mathcal{H}_c$
    	\Statex \hskip1.0em Gates: $e^{-iH\tau}\ (\tau>0)$ and controlled-Pauli gates on $\mathcal{H}_c\otimes\mathcal{H}$
    	\Statex \hrulefill
        \Statex{\textbf{Procedure:}}
        \State Compute $N:=  {\rm ceil} \left[\max\left(\frac{5\beta^2 {t}^2\Delta_H^2}{\epsilon}, \frac{5}{2}\beta t\Delta_H\right) \right]$ 
        \State{Initialize:}
        \Statex \hskip1.0em$\ket{\text{current}}\gets\ket{0}\otimes \ket{\psi}$
        \For{$m=1,\ldots ,N$}
        \State Randomly choose 
        \begin{itemize}
            \item $(\vec{v}, \vec{v}^{\prime})\in (\{0,1,2,3\}^n)^2$ with prob. $p_{\vec{v}, \vec{v}^{\prime}}^{(1)}:=\frac{1}{16^n}$
            \item $(\vec{u}, \vec{w})\in (\{0,1,2,3\}^n)^2$ with prob. $p_{\vec{u}, \vec{w}}^{(2)}:=\frac{2|\gamma_{\vec{u}, \vec{w}}|}{\beta}$
        \end{itemize}
    
        \State Prepare the gate sequence [with $j=(\vec{v},\vec{v}^{\prime},\vec{u},\vec{w})$]
        \begin{figure}[H]
            \begin{center}
            	\hspace{2.5em}
                \label{eq:V_fj}
                \includegraphics[width=7.5cm]{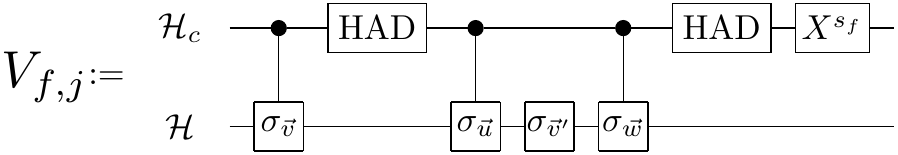}
            \end{center}
		\end{figure}
    	where  $s_f:=\frac{1-\mathrm{sgn}(\gamma_{\vec{u}, \vec{w}})}{2}$ (all gates other than $X^{s_f}$ are \indent independent of $f$) and $\rm HAD$ refers to the Hadamard \indent gate
    	
        \State $\ket{\text{current}}\gets V_{f,j}(I\otimes e^{-iHt\beta/N})V_{f,j}^{\dagger}\ket{\text{current}}$
        \EndFor
        \State Trace out $\mathcal{H}_c$ of $\ket{\text{current}}$
        \State {\textbf{Return}} $\ket{\text{current}}$
    \end{algorithmic}
\end{algorithm}

We assume that 
$f(I)\propto I$, which ensures that the resulting evolution $e^{-if(H)t}$ preserves the invariance under the global phase of $e^{-i H \tau}$. This class of $f$ covers all physically realizable linear transformations of $H$ as shown in Appendix C. 
In our setting, we are given the Pauli transfer matrices $\gamma$ \cite{chow2012universal} as in Eq.~(\ref{ptm}) of a hermitian-preserving linear map $f:\mathcal{L}(\mathcal{H})\to \mathcal{L}(\mathcal{H})$. 
Our algorithm simulates the Hamiltonian evolution $e^{-if(H)t}$ for any $t >0$ representing the time for the transformed Hamiltonian dynamics up to an error $\epsilon >0$ and variance $4\epsilon$ (see proof in Appendix A, which relies on more general results proven in Appendix B. A similar analysis of variance is obtained in probabilistic state synthesis \cite{Akibue}).

The runtime of our algorithm is upper-bounded as $O(\beta^2t^2\Delta_H^2n/\epsilon)$ in terms of $\beta := 2\sum_{\vec{w},\vec{u}}|\gamma_{\vec{w},\vec{u}}|$, which is a function of elements of the Pauli transfer matrix. The total evolution time of the input dynamics $e^{-iH\tau}$ is $\beta t$ which can be shown from step 3 and step 6 of the Algorithm 1.

\begin{figure}[t]
     \begin{center}
        \includegraphics[width=7cm]{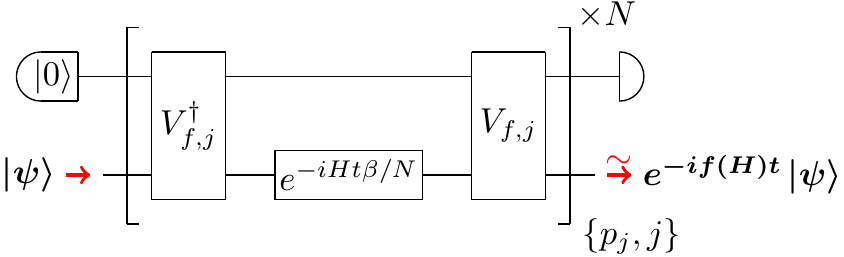}
     \end{center}
        \caption{A circuit representation of Algorithm 1 implementing the transformation $e^{-iH\tau} \mapsto e^{-if(H)t}$ for an arbitrary hermitian-preserving linear map $f:\mathcal{L}(\mathcal{H})\mapsto \mathcal{L}(\mathcal{H})$ satisfying $f(I)\propto I$. The unitary $e^{-i f(H)t}$ is simulated deterministically and approximately, for an arbitrary input state $\ket{\psi} \in \mathcal{H}$ and the auxiliary qubit initialized in the state $\ket{0} \in \mathcal{H}_c$. The number $N$ on the top-right of the bracket refers to the number of iterations while $t\beta/N$ is the Hamiltonian evolution time of each iteration. For each iteration, an index $j=(\vec{v},\vec{v}^{\prime},\vec{u},\vec{w})$ is randomly chosen from the probability distribution $p_j = p_{\vec{v},\vec{v}^{\prime}}^{(1)}p_{\vec{u},\vec{w}}^{(2)}$, to perform the $j$-dependent circuit inside the square brackets.}
        \label{overall_circuit}
\end{figure}

In Algorithm 1, the gate sequence $V_{f,j}$ is constructed only from controlled-Pauli gates, which are Clifford gates. The only element which may be non-Clifford is the black box dynamics $e^{-iH\tau}$. Dependence on the transformation $f$ is specified only through the probability distribution $p_{\vec{u}, \vec{w}}^{(2)}$ in choosing  $(\vec{u}, \vec{w})$ in Step 4 and through the gate $X^{s_f}$ in Step 5. The total runtime $O(\beta^2t^2\Delta_H^2n/\epsilon)$ is calculated by multiplying the number of iterations $N$ with the runtime $O(n)$ for implementing the controlled-Pauli gates in $V_{f,j}$ using CNOT gates and single-qubit Clifford gates. 
Note that $N$ is independent of $n$, even though the set of parameters $j\in (\{0,1,2,3\}^n)^4$ has exponentially many terms.
The procedure of Algorithm 1 is summarized in Figure \ref{overall_circuit}.

\begin{figure*}[t]
    \begin{center}
        \includegraphics[width=0.95\linewidth]{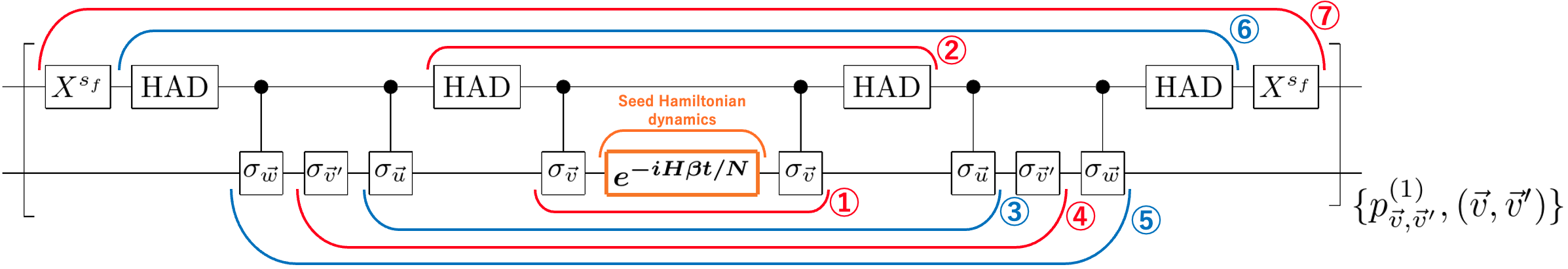}
     \end{center}
        \caption[.]{A description of how a seed Hamiltonian $H=\sum_{\vec{v}}c_{\vec{v}}\sigma_{\vec{v}}$ is transformed after each pair of gates in Algorithm 1, for a fixed choice of $\vec{u},\vec{w}$. The labels \textcircled{\scriptsize 1} to \textcircled{\scriptsize 7} correspond to the Processes defined in the text.
        }
 \label{fig:seq_of_transformations}
\end{figure*}

To understand how the gate sequence $V_{f,j}$ transforms the Hamiltonian at each iteration, Figure \ref{fig:seq_of_transformations} shows the explicit evolution of an arbitrary seed Hamiltonian $H=\sum_{\vec{v}}c_{\vec{v}}\sigma_{\vec{v}}$ after pre- and post-processing with each successive gate in the (random) sequence $V_{f,j}$ averaged over $\vec{v}$ and $\vec{v}^\prime$, namely, $\frac{1}{16^n} \sum_{\vec{v},\vec{v}^{\prime}}  V_{f,j} ( I \otimes  e^{-iH t\beta/N}) V_{f,j}^\dagger$.
For simplicity, $H$ is assumed to be traceless (any trace-full part is proportional to the identity and is therefore invariant under the overall transformation $f$, by construction).  
The gate sequence of $\frac{1}{16^n} \sum_{\vec{v},\vec{v}^{\prime}}V_{f,j} ( I \otimes  e^{-iH t\beta/N}) V_{f,j}^\dagger$ is constructed in a \textit{functional programming approach}, namely, by concatenations of a series of higher-order transformations, here called Processes \textcircled{\scriptsize 1} to \textcircled{\scriptsize 7}.
Each of these processes is designed to implement a Hamiltonian dynamics whose Hamiltonian is given by
        \begin{align}
             &I \otimes H = \left(
               \begin{array}{cc}
                    H&0\\
                    0&H\\
                \end{array}
            \right) \xrightarrow{\text{\textcircled{\scriptsize 1}} }
            \left(
                \begin{array}{cc}
                    H&0\\
                    0&0\\
                \end{array}
            \right) \xrightarrow{\text{\textcircled{\scriptsize 2}} }
            \left(
                \begin{array}{cc}
                    H&H\\
                    H&H\\
                \end{array}
            \right) \nonumber \\
            & \xrightarrow{\text{\textcircled{\scriptsize 3}} }
            \left(
                \begin{array}{cc}
                    H&H\sigma_{\vec{u}}\\
                    \sigma_{\vec{u}}H&\sigma_{\vec{u}}H\sigma_{\vec{u}}\\
                \end{array}
            \right) \xrightarrow{\text{\textcircled{\scriptsize 4}} } 
            c_{\vec{u}}
            \left(
                \begin{array}{cc}
                    0&I\\
                    I&0\\
                \end{array}
            \right) 
 \xrightarrow{\text{\textcircled{\scriptsize 5}} }
            c_{\vec{u}}
            \left(
                \begin{array}{cc}
                    0&\sigma_{\vec{w}}\\
                    \sigma_{\vec{w}}&0\\
                \end{array}
            \right)\nonumber\\ 
            &  \xrightarrow{\text{\textcircled{\scriptsize 6}} }
            c_{\vec{u}}
            \left(
                \begin{array}{cc}
                    \sigma_{\vec{w}}&0\\
                    0&-\sigma_{\vec{w}}\\
                \end{array}
            \right) \xrightarrow{\text{\textcircled{\scriptsize 7}} }
            \mathrm{sgn}(\gamma_{\vec{w},\vec{u}})c_{\vec{u}} 
            \left(
                \begin{array}{cc}
                    \sigma_{\vec{w}}&0\\
                    0&-\sigma_{\vec{w}}\\
                \end{array}
            \right).
            \nonumber
        \end{align} 

Applying the first controlled-$\sigma_{\vec{v}}$ gate before and after the seed Hamiltonian evolution $e^{-iH t \beta /N}$ with $\vec{v}$ chosen independently from the uniform distribution in each iteration but perfectly correlated between the pre- and post-processing within each iteration (Process \textcircled{\scriptsize 1}) implements Hamiltonian controllization  \cite{dong2019controlled}. That is, the effective evolution $({\tt ctrl}\sigma_{\vec{v}}) e^{-i(I\otimes H) t \beta /N} ({\tt ctrl}\sigma_{\vec{v}})$ averaged over $\vec{v}$ simulates a Hamiltonian of the form $H \oplus 0$.

Process \textcircled{\scriptsize 4} is based on the identity
\begin{align}
    &\frac{1}{4^n}\sum_{\vec{v}^{\prime}\in \{0,1,2,3\}^n}
    (I\otimes \sigma_{\vec{v}^{\prime}} )
    \left(
        \begin{array}{cc}
            H_{00}&H_{01}\\
            H_{01}^{\dagger}&H_{11}\\
        \end{array}
    \right)
    (I\otimes \sigma_{\vec{v}^{\prime}})
    \nonumber
    \\
    &\equiv \frac{1}{2^n}
    \left(
        \begin{array}{cc}
            \mathrm{tr}H_{00}&\mathrm{tr} H_{01}\\
            \mathrm{tr}H_{01}^{\dagger} & \mathrm{tr}H_{11}\\
        \end{array}
    \right)
    \otimes I,
\end{align}
where $H_{00},\ H_{01},\ H_{11}\in \mathcal{L}(\mathcal{H})$ are arbitrary operators, noting that for all $\vec{u}$, $\mathrm{tr}(H)=\mathrm{tr}(\sigma_{\vec{u}}H\sigma_{\vec{u}})=0$ and $\mathrm{tr}(\sigma_{\vec{u}}H)=\mathrm{tr}(H\sigma_{\vec{u}})=2^n c_{\vec{u}}$. The two gates $\sigma_{\vec{v}^{\prime}}$ are chosen independently from the uniform distribution at each iteration.

Algorithm 1 is universal in the sense that it transforms the dynamics of any seed Hamiltonian $H$ to that of the Hamiltonian $f(H)$ for any choice of a physically realizable linear transformation $f$, even if $H$ is only given as a black box. 
Therefore the algorithm is an instance of higher-order quantum transformations of Hamiltonian dynamics.
The algorithm makes use of a general approximation technique for simulating Hamiltonians of the form $g(H)=\sum_jh_jU_jHU_j^{\dagger}$, where $\{U_j\}_j$ is a set of unitaries, $\{h_j\}_j$ is a  set of positive numbers and $H$ is a seed Hamiltonian. This approximation is represented by the following circuit
\begin{figure}[H]
    \begin{equation}
        \includegraphics[width=7cm]{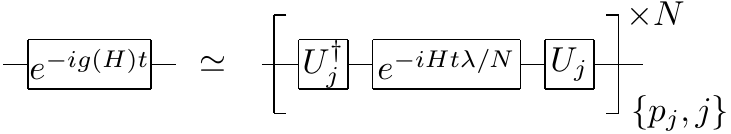}
        \label{uhuuhu}
    \end{equation}
\end{figure}
\noindent where $\lambda$ and $p_j$ are defined as $\lambda := \sum h_j$ and $p_j:=h_j/\lambda$. The approximation is based on the randomized Hamiltonian simulation of Ref.\ \cite{campbell2019random} and the identity $Ue^{-iHt}U^{\dagger} \equiv e^{-iUHU^{\dagger}t}$ for any unitary $U$, time $t > 0$ and hermitian operator $H$. This technique is also known as Hamiltonian reshaping \cite{huang2023learning}. Our algorithm can be seen as a special case of the approximation \eqref{uhuuhu} with $h_j = 2|\gamma_{\vec{u},\vec{w}}|  / 16^n$ and $\ U_j = V_{f,j}$, where the seed Hamiltonian has the form $I\otimes H$.

{\it Applications of the algorithm.---}We describe three applications of our algorithm: the negative time-evolution of Hamiltonian dynamics $e^{-iH\tau}\mapsto e^{iHt}\ (\tau,t>0)$, the time-reversal of Hamiltonian dynamics $e^{-iH\tau}\mapsto e^{-iH^Tt}\ (\tau,t>0)$, and a Hamiltonian single-parameter learning task of estimating one of the parameters represented by a Pauli coefficient $c_{\vec{v}}$ $(|c_{\vec{v}}|\leq 1\ ,\vec{v}\in \{0,1,2,3\}^n)$ of a Hamiltonian $H=\sum_{\vec{u}}c_{\vec{u}}\sigma_{\vec{u}}$ with Heisenberg-limited precision scaling using its dynamics $e^{-iH\tau}$ $(\tau >0)$.

In general, all three applications can be performed even if the dynamics $e^{-iH\tau}$ is given as a black box, apart from knowledge of $\Delta_H$. However, given the knowledge that $H$ belongs to a subspace of $\mathcal{L}(\mathcal{H})$ spanned by the set $\{\sigma_{\vec{v}}\}_{\vec{v}\in J}$ for some $J\subset {\{0,1,2,3\}^n}$, negative time-evolution and time-reversal can be performed in a runtime of $O[\mathrm{poly}(|J|)]$. This property is useful when the Hamiltonian is known to be $k$-local for some constant $k$, in which case $J=\left\{\vec{w}\ : ||\vec{w}||_0 \leq k \right\}$ satisfies $|J| \sim O(n^k)$, so that the overall runtime is polynomial in the system size $n$, based on the fact that $\Delta_H$ is also ${\rm poly}(n)$.

In quantum algorithms that make direct use of Hamiltonian dynamics, both the positive and negative time-evolution are often assumed to be readily accessible. For example, this is required in the recent Hamiltonian singular value transformation \cite{lloyd2021hamiltonian}.  However, in practice, a Hamiltonian evolution being native to a given hardware does not automatically guarantee that the same is true for the corresponding negative time-evolution. Therefore, the ability to efficiently simulate the negative time-evolution of any Hamiltonian given as a black box can decrease the resources required for such algorithms. On the more foundational side, given access to a black box Hamiltonian evolution, one might be interested in simulating the corresponding time-reversed evolution. For example, the evolution of an antiparticle can be described by the time-reversal of the corresponding particle evolution \cite{weinberg, Sakurai}.

The simulations of both negative time-evolution and time-reversal are performed by choosing the function $f$ as $f^{\rm neg}(H):=-H$ and $f^{\rm rev}(H):=H^T$, respectively, which are specified by 
\begin{align}
    \gamma^{\rm neg}_{\vec{w},\vec{u}} &:=-\delta_{\vec{w},\vec{u}}  
    \nonumber\\
    \gamma^{\rm rev}_{\vec{w},\vec{u}} &:=(-1)^{s_{\vec{w}}} \delta_{\vec{w},\vec{u}}, 
\end{align}
where $s_{(w_1,\ldots ,w_n)}:= |\{j\in \{1,\ldots ,n\} \mid w_j=2\}|$.
In the definition of $\gamma^{\rm rev}_{\vec{w},\vec{u}}$, the fact that $I^T=I$, $X^T=X$, $Y^T=-Y$, and $Z^T=Z$ are used. 

In both of these cases, $\beta = 2\sum_{\vec{w},\vec{u}}|\gamma_{\vec{w},\vec{u}}| = 2(4^n-1)$, thus the runtime $O(\beta ^2t^2\Delta_H^2 n/\epsilon)$ is exponential in $n$ in general. However, when $H$ is in a subspace of $\mathcal{L}(\mathcal{H})$ spanned by the set $\{\sigma_{\vec{v}}\}_{\vec{v}\in J}$ ,
we can define
\begin{align}
    \gamma^{\rm neg}_{\vec{w},\vec{u}}&:=
    \begin{cases}
        -\delta_{\vec{w},\vec{u}}&(\vec{u}\in J)\\
        0&(\text{otherwise})\\
    \end{cases}\\
    \gamma^{\rm rev}_{\vec{w},\vec{u}}&:=
    \begin{cases}
        (-1)^{s_{\vec{w}}} \delta_{\vec{w},\vec{u}}&(\vec{u}\in J)\\
        0&(\text{otherwise})\, ,\\
    \end{cases}
\end{align}
since $f(H)$ does not depend on values of $\gamma_{\vec{w},\vec{u}}$ for $\vec{u}\notin J$. 
In this case, $\beta = 2|J|$ so the runtime scales as  $O(|J|^2t^2\Delta_H^2 n/ \epsilon)$, which is $O[\mathrm{poly}(n)]$ for a realistic Hamiltonian whose number of terms $|J|$ is polynomial in the system size $n$.
For a general Hamiltonian linear transformation $f$,  if both the seed Hamiltonian and the transformed Hamiltonian have a polynomial number of terms in $n$, then the non-zero elements of $f$ can be truncated so that the runtime $O(\beta^2t^2\Delta_H^2n/\epsilon)$ has a polynomial dependence on $n$.

We note that the runtime scales as $t^2$, meaning that in order to perform the time-reversal or negative time-evolution by this algorithm, the dynamics is slowed down quadratically. An application of simulating the negative time-evolution to Hamiltonian block encoding \cite{lloyd2021hamiltonian} is described in Appendix D.

Finally, we consider an application of our algorithm to Hamiltonian single-parameter learning.  Estimation techniques of parameters of unknown Hamiltonians for Hamiltonian learning have many applications in quantum sensing \cite{de2005quantum}, analyzing properties of quantum many-body physics \cite{wiebe2014quantum}, and quantum device calibration \cite{boulant2003robust}. Recently, an estimation technique achieving the Heisenberg limit for the precision scaling in the estimation of parameters of a {\it low-interaction Hamiltonian} utilizing transformations of Hamiltonian dynamics has been proposed \cite{huang2023learning}. Our algorithm can be used to extend similar techniques to a more general class of $n$-qubit Hamiltonians.

Our estimation algorithm consists of two steps. The first step simulates $e^{-if_{\vec{v}}(H_0)t}$ $(t>0)$ using the Hamiltonian dynamics $e^{-iH\tau}$ $(\tau >0)$ where $\vec{v}$ specifies $c_{\vec{v}}$ that we want to estimate and $f_{\vec{v}}$ is a hermitian-preserving linear map chosen as 
$f_{\vec{v}}(H)=c_{\vec{v}}Y\otimes I\otimes \cdots \otimes I$.
This function $f_{\vec{v}}$ ``filters'' to keep only the coefficient $c_{\vec{v}}$ and changes all other coefficients to be zero, and then sends the coefficient $c_{\vec{v}}$ to the coefficient of $Y\otimes I\otimes \cdots \otimes I$, which is chosen for the convenience of the second step. The corresponding $\gamma$ is given by $\gamma_{\vec{w}\vec{u}}:=\delta_{\vec{w},(2,0,\ldots ,0)}\delta_{\vec{u},\vec{v}}$.   
The second step performs robust phase estimation \cite{kimmel2015robust} using $e^{-if_{\vec{v}}(H)t}$ similarly to the technique in \cite{huang2023learning} to obtain an estimate for $c_{\vec{v}}$, by measuring only the first qubit in our algorithm. The total evolution time is $O((\log \delta )/\epsilon)$ where $\epsilon$ is precision and $\delta$ is the failure probability, which achieves the Heisenberg-limited precision scaling. The detailed procedure and analysis of the total evolution time are given in Appendix E.

For parameter estimation of low-interaction Hamiltonians, the method of \cite{huang2023learning} can perform the full-parameter estimation in a single run with total evolution time $O((\log \delta )/\epsilon)$, while our method requires polynomially longer total evolution time as we need to repeat the single parameter estimation for every parameter to perform the same task. However, the method of \cite{huang2023learning} requires exponential total evolution time for estimating a high-interaction coefficient (a coefficient of a $k$-local Hamiltonian term with $k=O(n)$), while our algorithm requires the same total evolution time for any coefficient. Therefore, our algorithm is suitable for estimating a single parameter of non-local Hamiltonians.

{\it Summary and outlook.---}
We presented a universal algorithm that can simulate any linear physically realizable hermitian-preserving transformation of any Hamiltonian dynamics given as a black box. 
Our algorithm requires only a finite number of calls to the black box Hamiltonian dynamics and random pairs of correlated controlled-Pauli gates. We showed how our algorithm can simulate both the time-reversal and negative time-evolution of any unknown Hamiltonian dynamics, as well as an application to Hamiltonian single-parameter learning, efficiently estimating a single parameter of a multi-parameter Hamiltonian. 

In our algorithm, the probability distribution for choosing multiple gates at different time steps are correlated in the sense that the gate $V_{f,j}$ is always used together with its adjoint $V_{f,j}^{\dagger}$, and the probabilities for picking its component controlled-Pauli gates are correlated via a joint probability distribution. This algorithm demonstrates how multiply correlated randomness can be leveraged to construct unitary operators without introducing decoherence.
Our algorithm is a starting point for the emerging field of black box Hamiltonian simulation. One possible future direction is to extend higher-order quantum transformations of Hamiltonian dynamics to Hamiltonian transformations beyond hermitian-preserving linear transformations.

We would like to thank Seiseki Akibue, Mile Gu, Toshinori Itoko, Antonio Mezzacapo, Kunal Sharma, Philip Taranto, and Satoshi Yoshida for fruitful discussions.  This work was supported by the MEXT Quantum Leap Flagship Program (MEXT QLEAP) Grant No.\ JPMXS0118069605, the Japan Society for the Promotion of Science (JSPS) KAKENHI Grants No.\ 21H03394 and  No.\ 18K13467, and partly by IBM-UTokyo lab. Research at Perimeter Institute is supported in part by the Government of Canada through the Department of Innovation, Science and Economic Development and by the Province of Ontario through the Ministry of Colleges and Universities.

\appendix

\clearpage

\section{Appendix A: Error and variance analysis on Algorithm 1} \label{app:proof}

In this section, we prove a theorem (Theorem 1) evaluating the error and variance of Algorithm 1 in simulating $e^{-if(H)t}$.  We use the same notation and symbols appearing in Algorithm 1 in this appendix and the following appendices.

For evaluating the error of the simulated quantum operation from the target quantum operation, we use the diamond norm $\|  \mathcal{E} \|_{\diamond}$ of a quantum operation  $ \mathcal{E}: \mathcal{L}(\mathcal{H})\to \mathcal{L}(\mathcal{H})$ defined as \cite{watrous2018theory}
        \begin{align}
            \|  \mathcal{E}\|_{\diamond}&:=
            \underset{A ;\|A \|_1=1}{\mathrm{sup}}
            \| ( \mathcal{E}\otimes \mathcal{I})(A)
            \|_1
       \end{align} 
where the identity operation $\mathcal{I}$ acts on a Hilbert space isomorphic to $\mathcal{H}$, $A$ is an arbitrary linear operator on $\mathcal{H}\otimes \mathcal{H}$, and $\|A \|_1$ is the trace norm defined by $\|A \|_1:= \mathrm{tr}(\sqrt{A ^{\dagger}A})$.

\begin{Theorem}
	{\color{white}.} \newline
    \begin{enumerate}
        \item The approximation error of Algorithm 1 is given by
        \begin{align}
            \frac{1}{2}\| \mathcal{F}-{ \mathcal{F}}_{\mathrm{approx}}\|_{\diamond}
            \label{our_error}\leq \epsilon \,,
        \end{align}
        where $\mathcal{F}$ is a map defined by $\mathcal{F}(\rho):= e^{-if(H)t}\rho e^{if(H)t}$ and ${ \mathcal{F}}_{\mathrm{approx}}$ is the quantum operation averaged over all random instances of operations performed in Algorithm 1. 
        \item The variance of Algorithm 1 is given by
        \begin{align}
            \underset{\substack{ 
            \mathrm{dim}(\mathcal{H}^{\prime}) \\
            \ket{\psi}\in \mathcal{H}\otimes \mathcal{H}^{\prime}\\
            \|\ket{\psi}\|=1 }}{\mathrm{sup}}
            \sum_{\vec{j}} p_{\vec{j}} \|
            ({\mathcal{F} \otimes \mathcal{I}}_{\mathcal{H}^{\prime}})(\ket{\psi}\bra{\psi}) &-
            \nonumber\\
            ({\mathcal{F}}_{\vec{j}}\otimes { \mathcal{I}}_{\mathcal{H}^{\prime}})(\ket{\psi}\bra{\psi})
            \|_1^2
            &\leq 4\epsilon
            \,,
        \end{align}
        where $\vec{j}=(j_1,\ldots ,j_{N})$ refers to the set of all indices $j$ chosen in $N$ iterations, $p_{\vec{j}}$ is the probability that $\vec{j}$ is chosen, and ${ \mathcal{F}}_{\vec{j}}$ is the unitary performed when $\vec{j}$ is chosen.
    \end{enumerate}
\end{Theorem}

Without loss of generality, we can limit the proof of Theorem 1 to the case where $\|H_0\|_{{\rm op}}<1$, where $H_0:=H-({\rm tr}H/2^n) I$ is the traceless part of $H$. When $\|H\|_{{\rm op}}>1$, then $\Delta_H \geq \max_{j,k}|E_j-E_k|>\|H_0\|_{{\rm op}}$ will also be greater than 1, where where $\{E_k\}_k$ is the set of eigenvalues of $H$. By noticing that the procedure of Algorithm 1 for a Hamiltonian $H$ and time $t$ is the same as that for a Hamiltonian $H':=H/\Delta_H$ (whose traceless part has operator norm of at most 1) and time $t':=t\Delta_H$, we can always assume that the traceless part of the seed Hamiltonian has operator norm at most 1.
Before presenting the sketch of the proof, we describe the error and variance of a simulation technique shown in the circuit of Fig. 4, which is a key element in our algorithm.  Similar randomization techniques are also used in \cite{campbell2019random} (QDrift) and \cite{huang2023learning} (Hamiltonian learning).  The simulation error is evaluated based on \cite{campbell2019random}.  In addition, the variance is also evaluated for our algorithm.

\begin{lem}
    The quantum operation represented by the following circuit
    \begin{figure}[H]
        \begin{center}
            \includegraphics[width=6cm]{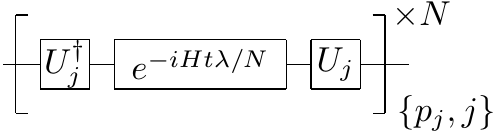}
        \end{center}
    \end{figure}
\noindent where $H$ is a Hamiltonian normalized as $\|H_0\|_{{\rm op}}\leq 1$, where $H_0:=H-({\rm tr}H/2^n) I$ is the traceless part of $H$, $\lambda>0$, $t>0$ are constants, $U_j$ is a unitary applied with probability $p_j$, and $N$ is the iteration number given by $N= {\rm ceil}[\mathrm{max}(\frac{5\lambda ^2 t^2}{\epsilon}, \frac{5}{2} \lambda t)]$ for $\epsilon>0$,
simulates the dynamics $e^{-ig(H)t}$ of a transformed Hamiltonian $g(H)$ defined as 
    \begin{align}
        g(H):=\sum_{j}h_jU_jHU_j^{\dagger}
        \label{uhuhu}
    \end{align}
with $h_j = \lambda p_j >0$
    in an error smaller than or equal to $\epsilon>0$ and the variance smaller than or equal to $4\epsilon$. 
    
    Here, the error and the variance of the simulation are defined as
    \begin{align}
        &\mathrm{Error:} \,
        \frac{1}{2}\|{\mathcal{G}}-{\mathcal{G}}_{\mathrm{approx}}\|_{\diamond}
        \label{error_def}\\
        &\mathrm{Variance:}
        \nonumber\\
        &\underset{\substack{ \mathrm{dim}(\mathcal{H}^{\prime}) \\ \ket{\psi}\in \mathcal{H}\otimes \mathcal{H}^{\prime}\\ \ket{\psi};\|\ket{\psi}\|=1 }}{\mathrm{sup}}
            \sum_{\vec{j}} q_{\vec{j}} \|
            (\mathcal{G}\otimes \mathcal{I})(\ket{\psi}\bra{\psi}) -
            ({\mathcal{G}}_{\vec{j}}\otimes \mathcal{I})(\ket{\psi}\bra{\psi})
            \|_1^2 \, ,
        \label{variance_def}
    \end{align}
respectively, where $\mathcal{G}$ is defined by $$\mathcal{G}(\rho ):=e^{-ig(H)t}\rho e^{ig(H)t}$$ for a density operator $\rho$, ${\mathcal{G}}_{\mathrm{approx}}$ is the averaged quantum operation simulated in the above circuit, $\vec{j}=(j_1,\ldots ,j_{N})$ refers to the set of all the indices $j$ chosen in $N$ iterations, $q_{\vec{j}}$ is the probability $\vec{j}$ is chosen, ${\mathcal{G}}_{\vec{j}}$ is the unitary operation performed when $\vec{j}$ is chosen, and the identity operation $\mathcal{I}$ belongs to a Hilbert space $\mathcal{H}^{\prime}$ with an arbitrary finite dimension.
\end{lem}

\Proof \\
Both $\mathcal{G}$ and $\mathcal{G}_{{\rm approx}}$ stay unchanged when the seed Hamiltonian $H$ is changed to $H_0$ since the two dynamics $e^{-iHt}$ and $e^{-iH_0t}$ for the same time $t$  differ only by a global phase. Therefore, giving a proof for $H=H_0$ is sufficient.\\
        \indent \textit{Error:}
        The set $\{H_j\}$ of hermitian operators for $H_j:=U_j H_0U_j^{\dagger}$ satisfies $\|H_j\|_{\mathrm{op}}=1$, thus the protocol given by
        \begin{figure}[H]
            \begin{center}
                \includegraphics[width=4cm]{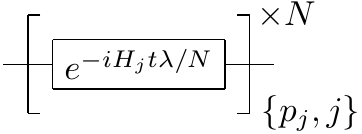}
            \end{center}
        \end{figure}
        \noindent with $h_j>0$, $N:={\rm ceil}[\mathrm{max}(\frac{5\lambda ^2 t^2}{\epsilon}, \frac{5}{2} \lambda t)]$, and $\lambda :=\sum_{j}h_j$ simulates $\rho \mapsto e^{-it\sum_jh_jH_j}\rho e^{it\sum_jh_jH_j}$ in an error smaller than $ \epsilon$. 
        Here, the error is measured in terms of Eq.\ (\ref{error_def}) with $g(H):=\sum_{j}h_j H_j$.
        This is proved using the fact that the error is smaller than or equal to $\frac{2\lambda ^2 t^2}{N}e^{2 \lambda t/N}$ as shown in \cite{campbell2019random} and $\frac{2\lambda ^2 t^2}{N}e^{2 \lambda t/N}\leq \frac{2\epsilon}{5}e^{4/5}\leq \epsilon$.
        
     \textit{Variance:} The variance is proven by using the upper bound shown by Theorem 3 of Appendix B given by
        \begin{align}
            \frac{1}{2}\|G-G_{\mathrm{approx}}\|_{\diamond}\leq \epsilon.
        \end{align}
    \qed

We refer to linear maps $g$ expressed by Eq.\ (\ref{uhuhu}) as linear maps in Class T. 
\begin{de}
    Class T of linear maps $\mathcal{L}(\mathcal{H})\mapsto \mathcal{L}(\mathcal{H})$ is defined as a set of linear maps $g:\mathcal{L}(\mathcal{H})\mapsto \mathcal{L}(\mathcal{H})$ which can be specified by a set $\{(h_j,U_j)\}_j$ of positive numbers $h_j$ and unitaries $U_j$ on $\mathcal{H}$ through the equation
    \begin{align}
        g(H)=\sum_{j}h_jU_jHU_j^{\dagger},
        \nonumber
    \end{align}
    where $H$ is an hermitian operator $H\in \mathcal{L}(\mathcal{H})$.
\end{de}
Class T is closed to concatenation, namely, the following lemma holds.
\begin{lem}
        A finite concatenations $g^{(K)}\circ \cdots \circ g^{(1)}$ $(K\in \mathbb{Z}_{>0})$ of Class T transformations $g^{(1)},\ldots ,g^{(K)}$ also belong to class T. In particular, if $g^{(k)}$ for $k\in \{1,\ldots ,K\}$  is specified by $\{ (h^{(k)}_{j^{(k)}}, U^{(k)}_{j^{(k)}}) \}_{j^{(k)}}$, then $g^{(k)}\circ \cdots \circ g^{(1)}$ is specified by 
        $\{ ( h^{(K)}_{j^{(K)}}\cdots h^{(1)}_{j^{(1)}}, U^{(K)}_{j^{(K)}}\cdots U^{(1)}_{j^{(1)}} ) \}_{(j^{(1)},\ldots ,j^{(K)})}$. Here, the index sets $j^{(k)}$ can be individually chosen for $k\in \{1,\ldots ,K\}$.
\end{lem}

\Proof
    It can be proved from the equation given by
    \begin{align}
        &(g^{(K)}\circ \cdots \circ g^{(1)})(H)=
        \nonumber\\
        &\sum_{(j^{(1)},\ldots ,j^{(K)})}
        h^{(K)}_{j^{(K)}}\cdots h^{(1)}_{j^{(1)}}
        U^{(K)}_{j^{(K)}}\cdots U^{(1)}_{j^{(1)}}
        H
        U^{(1)}_{j^{(1)}}\cdots U^{(K)}_{j^{(K)}}.
        \nonumber
    \end{align}
    \qed

In the following lemma, $f,n,\gamma_{\vec{w},\vec{u}},\beta,j,s_f,V_{f,j}$ follow the  definitions in the description of Algorithm 1; $p_j := p_{\vec{v},\vec{v}^{\prime}}^{(1)}p_{\vec{u},\vec{w}}^{(2)}$. 
\begin{lem}\label{le::gtotal}
    \begin{enumerate}
        \item 
        The Class T transformation $g_{\mathrm{total}}:\mathcal{L}(\mathcal{H}_c\otimes \mathcal{H})\mapsto \mathcal{L}(\mathcal{H}_c\otimes \mathcal{H})$ specified by $\{(h_j, V_{f,j})\}_j$ for $h_j:=\beta p_j$ is expressed as a sum of concatenations of Class T transformations as
        \begin{align}
            g_{\mathrm{total}}:=\sum_{\vec{w},\vec{u}}|\gamma_{\vec{w},\vec{u}}|
            g^{(7)}_{f,\vec{w},\vec{u}}\circ
            g^{(6)}\circ
            g^{(5)}_{\vec{w}}\circ
            g^{(4)}\circ
            g^{(3)}_{\vec{u}}\circ
            g^{(2)}\circ
            g^{(1)},
            \label{concat_sum}
        \end{align}
        where $g^{(1)},\ g^{(2)},\ g^{(3)}_{\vec{u}},\ g^{(4)},\ g^{(5)}_{\vec{w}},\ g^{(6)},\ g^{(7)}_{f,\vec{w},\vec{u}}$ are defined as
        \begin{align}
            g^{(1)}(\tilde{H}) &:=
            \frac{1}{4^n}\sum_{\vec{v}}
            \left(
            \begin{array}{cc}
            I&0\\
            0&\sigma_{\vec{v}}\\
            \end{array}
            \right)
            \tilde{H}
            \left(
            \begin{array}{cc}
            I&0\\
            0&\sigma_{\vec{v}}\\
            \end{array}
            \right)
            \nonumber \\
            g^{(2)}(\tilde{H}) &:=
            2(\text{HAD}\otimes I)
            \tilde{H}
            (\text{HAD}\otimes I)
            \nonumber \\
            g^{(3)}_{\vec{u}}(\tilde{H}) &:=
            \left(
            \begin{array}{cc}
            I&0\\
            0&\sigma_{\vec{u}}\\
            \end{array}
            \right)
            \tilde{H}
            \left(
            \begin{array}{cc}
            I&0\\
            0&\sigma_{\vec{u}}\\
            \end{array}
            \right)
            \nonumber \\
            g^{(4)}(\tilde{H}) &:=
            \frac{1}{4^n}
            \sum_{\vec{v}^{\prime}}
            (I\otimes \sigma_{\vec{v}^{\prime}})
            \tilde{H}
            (I\otimes \sigma_{\vec{v}^{\prime}})
            \nonumber \\
            g^{(5)}_{\vec{w}}(\tilde{H})&:=
            \left(
            \begin{array}{cc}
            I&0\\
            0&\sigma_{\vec{w}}\\
            \end{array}
            \right)
            \tilde{H}
            \left(
            \begin{array}{cc}
            I&0\\
            0&\sigma_{\vec{w}}\\
            \end{array}
            \right)
            \nonumber \\
            g^{(6)}(\tilde{H}) &:=
            (\text{HAD}\otimes I)
            \tilde{H}
            (\text{HAD}\otimes I)
            \nonumber \\
            g^{(7)}_{f,\vec{w}, \vec{u}}(\tilde{H}) &:=
            (X^{s_f}\otimes I)
            \tilde{H}
            (X^{s_f}\otimes I)
            \nonumber
        \end{align}
        for an input operator $\tilde{H}\in \mathcal{L}(\mathcal{H}_c\otimes \mathcal{H})$.
        \item 
        The transformation $g_{\mathrm{total}}$ transforms an input operator $I\otimes H$ for a hermitian operator $H\in \mathcal{L}(\mathcal{H})$ as
        \begin{align}
            g_{\mathrm{total}}(I\otimes H)=
            \left(
            \begin{array}{cc}
            f(H)&0\\
            0&-f(H)\\
            \end{array} 
            \right)
            + \mathrm{const.}\times 
            \left(
            \begin{array}{cc}
            I&0\\
            0&I\\
            \end{array} 
            \right).
           \label{lemma4finaleq} 
        \end{align}
    \end{enumerate}
\end{lem}

Note that the quantum operation performed in step 3 to step 7 of Algorithm 1 approximates the unitary $e^{-ig_{\mathrm{total}}(I\otimes H)t}$. Indeed, the circuit shown in Lemma 2 for $h_j$ defined above and $V_{f,j}$ defined inside Algorithm 1 coincides with the procedures in step 3 to step 7 due to $\sum_j h_j=\beta$ and $h_j/\beta = p_j$.
The functions $g^{(1)},\ g^{(2)},\ g^{(3)}_{\vec{u}},\ g^{(4)},\ g^{(5)}_{\vec{w}},\ g^{(6)},\ g^{(7)}_{f,\vec{w},\vec{u}}$ correspond to the transformation of the effective Hamiltonian by Processes \text{\textcircled{\scriptsize 1}} to \text{\textcircled{\scriptsize 7}} of Fig. 2.
 
\Proof
        
1. Class T transformations $g^{(1)}$, $g^{(2)}$, $g^{(3)}_{\vec{u}}$, $g^{(4)}$, $g^{(5)}_{\vec{w}}$, $g^{(6)}$, $g^{(7)}_{f,\vec{w},\vec{u}}$ are specified by 
        $\{ (1/4^n,\mathrm{Controlled-}\sigma_{\vec{v}} ) \}_{\vec{v}}$, 
        $\{(2, \mathrm{HAD}\otimes I)\}$, 
        $\{ (1, \mathrm{Controlled-}\sigma_{\vec{u}} )\}$, 
        $\{ (1/4^n, I\otimes \sigma_{\vec{v}^{\prime}}) \}_{\vec{v}^{\prime}}$, 
        $\{ (1, \mathrm{Controlled-}\sigma_{\vec{w}}) \}$,
        $\{(1, \mathrm{HAD}\otimes I)\}$, 
        $\{ (1, X^{s_f}\otimes I) \}$, thus the right hand side of Eq.\ (\ref{concat_sum}) consists of Class T transformations specified by
        \begin{align}
            &\{|\gamma_{\vec{w},\vec{u}}|\cdot \frac{1}{4^n}\cdot 2\cdot \frac{1}{4^n}, 
            (X^{s_f}\otimes I)\circ 
            (\mathrm{HAD}\otimes I)
            \nonumber\\
            &\circ 
            (\mathrm{Controlled-}\sigma_{\vec{w}})
            \circ(I\otimes \sigma_{\vec{v}^{\prime}})\circ
            (\mathrm{Controlled-}\sigma_{\vec{u}})
            \nonumber\\
            &\circ
            (\mathrm{HAD}\otimes I)\circ
            (\mathrm{Controlled-}\sigma_{\vec{v}})
            \}_{\vec{v},\vec{v}^{\prime},\vec{w},\vec{u}}
            \nonumber\\
            &=\{ (h_j, V_{f,j})\}_j
            \nonumber
        \end{align}
        
2. We describe how the effective Hamiltonians are transformed in each Class T transformation according to Processes \text{\textcircled{\scriptsize 1}} to \text{\textcircled{\scriptsize 7}}. By defining
        \begin{align} 
        g_{\textrm{total}_{\vec{w},\vec{u}}} := (g^{(7)}_{f,\vec{w},\vec{u}}\circ
            g^{(6)}\circ
            g^{(5)}_{\vec{w}}\circ
            g^{(4)}\circ
            g^{(3)}_{\vec{u}}\circ
            g^{(2)}\circ
            g^{(1)}),
        \end{align}
and using $f(H)=\sum_{\vec{w},\vec{u}}\gamma_{\vec{w},\vec{u}}c_{\vec{u}}\sigma_{\vec{w}}$, Eq.\ (\ref{lemma4finaleq}) is rewritten as
        \begin{align} 
        g_\textrm{total}(I \otimes H) &=  \sum_{\vec{w},\vec{u}}  
        |\gamma_{\vec{w},\vec{u}}|
        g_{\textrm{total}_{\vec{w},\vec{u}}} (I \otimes H) \nonumber \\
        & = \sum_{\vec{w},\vec{u}} \gamma_{\vec{w},\vec{u}}c_{\vec{u}}
            \left(
            \begin{array}{cc}
            \sigma_{\vec{w}}&0\\
            0& - \sigma_{\vec{w}}\\
            \end{array} 
            \right) \nonumber \\
            &+ \mathrm{const.} \times 
            \left(
            \begin{array}{cc}
            I&0\\
            0&I\\
            \end{array} 
            \right).
        \end{align}        
Therefore it is sufficient to prove 
        \begin{align}
            &g_{\textrm{total}_{\vec{w},\vec{u}}} (I \otimes H)
            \nonumber\\
            &=
            \mathrm{sgn}(\gamma_{\vec{w},\vec{u}})c_{\vec{u}}
            \left(
                \begin{array}{cc}
                    \sigma_{\vec{w}}&0\\
                    0&-\sigma_{\vec{w}}\\
                \end{array}
            \right) +\mathrm{const.}\times (I\otimes I) 
            \label{sufficient}
        \end{align}
        for a seed Hamiltonian $H=\sum_{\vec{u}}c_{\vec{u}}\sigma_{\vec{u}}$. 
        Using $\alpha $ defined by $H=H_0+\alpha I$ where $H_0$ is the traceless part of $H$ ($\mathrm{tr}(H_0)=0$), Eq. (\ref{sufficient}) is proved by calculating the effective Hamiltonian after each process as 
        \begin{align}
            g^{(1)}\left(
                \left(
                    \begin{array}{cc}
                        H&0\\
                        0&H\\
                    \end{array}
                \right)
            \right) =
            \left(
                \begin{array}{cc}
                    H_0&0\\
                    0&0\\
                \end{array}
            \right)
            +\alpha
            \left(
                \begin{array}{cc}
                    I&0\\
                    0&I\\
                \end{array}
            \right) ,
            \nonumber
        \end{align}
        \begin{align}
            &g^{(2)}\left(
                \left(
                \begin{array}{cc}
                    H_0&0\\
                    0&0\\
                \end{array}
            \right)
            +\alpha
            \left(
                \begin{array}{cc}
                    I&0\\
                    0&I\\
                \end{array}
            \right)
            \right) 
            \nonumber\\
            &=
                \left(
                    \begin{array}{cc}
                        H_0&H_0\\
                        H_0&H_0\\
                    \end{array}
                \right)
                +2 \alpha
                \left(
                    \begin{array}{cc}
                        I&0\\
                        0&I\\
                    \end{array}
                \right) ,
            \nonumber
        \end{align}
        \begin{align}
            &g^{(3)}_{\vec{u}}\left(
                \left(
                    \begin{array}{cc}
                        H_0&H_0\\
                        H_0&H_0\\
                    \end{array}
                \right)
                +2 \alpha
                \left(
                    \begin{array}{cc}
                        I&0\\
                        0&I\\
                    \end{array}
                \right)
            \right) 
            \nonumber\\
            &=
            \left(
                \begin{array}{cc}
                    H_0&H_0\sigma_{\vec{u}}\\
                    \sigma_{\vec{u}}H_0&\sigma_{\vec{u}}H_0\sigma_{\vec{u}}\\
                \end{array}
            \right)
            + 2\alpha
            \left(
                \begin{array}{cc}
                    I&0\\
                    0&I\\
                \end{array}
            \right) ,
            \nonumber
        \end{align}
        \begin{align}
            &g^{(4)}\left(
                \left(
                    \begin{array}{cc}
                        H_0&H_0\sigma_{\vec{u}}\\
                        \sigma_{\vec{u}}H_0&\sigma_{\vec{u}}H_0\sigma_{\vec{u}}\\
                    \end{array}
                \right)
                + 2\alpha
                \left(
                    \begin{array}{cc}
                        I&0\\
                        0&I\\
                    \end{array}
                \right)
            \right) 
            \nonumber\\
            &= 
                \left(
                    \begin{array}{cc}
                        0&c_{\vec{u}}I\\
                        c_{\vec{u}}I&0\\
                    \end{array}
                \right)
                + 2\alpha
                \left(
                    \begin{array}{cc}
                        I&0\\
                        0&I\\
                    \end{array}
                \right) ,
            \nonumber
        \end{align}
        \begin{align}
            &g^{(5)}_{\vec{w}}\left(
                \left(
                    \begin{array}{cc}
                        0&c_{\vec{u}}I\\
                        c_{\vec{u}}I&0\\
                    \end{array}
                \right)
                + 2\alpha
                \left(
                    \begin{array}{cc}
                        I&0\\
                        0&I\\
                    \end{array}
                \right)
            \right) 
            \nonumber\\
            &= c_{\vec{u}}
            \left(
                \begin{array}{cc}
                    0&\sigma_{\vec{w}}\\
                    \sigma_{\vec{w}}&0\\
                \end{array}
            \right)
            +2\alpha
            \left(
                \begin{array}{cc}
                    I&0\\
                    0&I\\
                \end{array}
            \right) ,
            \nonumber
        \end{align}
        \begin{align}
            &g^{(6)}\left(
                c_{\vec{u}}
            \left(
                \begin{array}{cc}
                    0&\sigma_{\vec{w}}\\
                    \sigma_{\vec{w}}&0\\
                \end{array}
            \right)
            +2\alpha
            \left(
                \begin{array}{cc}
                    I&0\\
                    0&I\\
                \end{array}
            \right)
            \right) 
            \nonumber\\
            &= c_{\vec{u}}
            \left(
                \begin{array}{cc}
                    \sigma_{\vec{w}}&0\\
                    0&-\sigma_{\vec{w}}\\
                \end{array}
            \right)
            +2\alpha
            \left(
                \begin{array}{cc}
                    I&0\\
                    0&I\\
                \end{array}
            \right) ,
            \nonumber
        \end{align}
        \begin{align}
            &g^{(7)}_{f,\vec{w},\vec{u}}\left(
                c_{\vec{u}}
            \left(
                \begin{array}{cc}
                    \sigma_{\vec{w}}&0\\
                    0&-\sigma_{\vec{w}}\\
                \end{array}
            \right)
            +2\alpha
            \left(
                \begin{array}{cc}
                    I&0\\
                    0&I\\
                \end{array}
            \right)
            \right) 
            \nonumber\\
            &=  \mathrm{sgn}(\gamma_{\vec{w},\vec{u}})c_{\vec{u}}
            \left(
                \begin{array}{cc}
                    \sigma_{\vec{w}}&0\\
                    0&-\sigma_{\vec{w}}\\
                \end{array}
            \right)
            +2\alpha
            \left(
                \begin{array}{cc}
                    I&0\\
                    0&I\\
                \end{array}
            \right),
            \nonumber
        \end{align}
        as the final effective Hamiltonian coincides with the right-hand side of Eq. (\ref{sufficient}).
The transformations in $g^{(1)}$ and $g^{(4)}$ are obtained due to the equality 
\begin{align}
    \frac{1}{4^n}\sum_{\vec{u}}\sigma_{\vec{u}}M\sigma_{\vec{u}}=
    \frac{\mathrm{tr}(M)}{2^n}I ,
    \nonumber
\end{align}
where $M\in \mathcal{L}(\mathcal{H})$ is an arbitrary linear operator,
which is proved using the fact that $\{\sigma_{\vec{u}} \}_{\vec{u}}$ gives a basis of $\mathcal{L}(\mathcal{H})$, $\mathrm{tr}(\sigma_{\vec{u}})/2^n = \delta_{\vec{u},(0,\ldots ,0)}$, and
\begin{align}
    \frac{1}{4^N}\sum_{\vec{v}^{\prime}}\sigma_{\vec{v}^{\prime}}
    \sigma_{\vec{v}}\sigma_{\vec{v}^{\prime}}=
    \begin{cases}
        \sigma_{\vec{0}}=I&(\vec{v}=\vec{0})\\
        0&(\text{otherwise})\\
    \end{cases}.
    \nonumber
\end{align}
    \qed

\textbf{Proof of Theorem 1.}
As we mentioned above, we limit without loss of generality the proof of Theorem 1 to the case where $\|H_0\|_{{\rm op}}<1$, where $H_0:=H-({\rm tr}H/2^n) I$ is the traceless part of $H$.
\begin{description}
    \item [Error] 
    We denote three types of quantum operations, $\mathcal{E}^{(1)}:\mathcal{L}(\mathcal{H})\to \mathcal{L}(\mathcal{H}_c\otimes\mathcal{H})$, $\mathcal{E}^{(2)}:\mathcal{L}(\mathcal{H}_c\otimes\mathcal{H})\to \mathcal{L}(\mathcal{H}_c\otimes\mathcal{H})$, $\mathcal{E}^{(3)}:\mathcal{L}(\mathcal{H}_c\otimes\mathcal{H})\to \mathcal{L}(\mathcal{H})$ depending on the domain and range of the operations. In Algorithm A, $\mathcal{E}^{(1)}(\rho )=\ket{0}\bra{0} \otimes \rho$ is performed in step 2, $\mathcal{E}^{(2)}$ is performed in step 3-7, and $\mathcal{E}^{(3)}(\rho^{\prime} )=\mathrm{tr}_{\mathcal{H}_c}
        (\rho ^{\prime})$ is performed in step 8 where $\rho\in \mathcal{L}(\mathcal{H})$ and $\rho ^{\prime}\in \mathcal{L}(\mathcal{H}_c\otimes\mathcal{H})$ are density operators.  
        
   For $\mathcal{F}$ representing the dynamics of transformed Hamiltonian defined by $$\mathcal{F}(\rho ^{\prime})=e^{-ig_{\mathrm{total}}(I\otimes H)t}\rho ^{\prime} e^{ig_{\mathrm{total}}(I\otimes H)t}$$ where $g_{\mathrm{total}}$ is the linear transformation defined in the first statement of Lemma \ref{le::gtotal}, it can be shown using Lemma 2 and Lemma \ref{le::gtotal} that
    \begin{align}
        \frac{1}{2}\|\mathcal{F}-\mathcal{E}^{(2)}\|_{\diamond}
        \leq \epsilon .
        \nonumber
    \end{align}
    Therefore, Algorithm 1 approximates $\mathcal{E}^{(3)}\circ \mathcal{F}\circ \mathcal{E}^{(1)}$ in an error bounded above as
    \begin{align}
        &\frac{1}{2}\|\mathcal{E}^{(3)}\circ \mathcal{F}\circ \mathcal{E}^{(1)}-
        \mathcal{E}^{(3)}\circ \mathcal{E}^{(2)}\circ \mathcal{E}^{(1)}\|_{\diamond}
        \nonumber\\
        &\leq 
        \frac{1}{2}(
        \|\mathcal{E}^{(3)}-
        \mathcal{E}^{(3)}\|_{\diamond}+
        \|\mathcal{F}-
        \mathcal{E}^{(2)}\|_{\diamond}+
        \|\mathcal{E}^{(1)}-
        \mathcal{E}^{(1)}\|_{\diamond})
        \leq \epsilon ,
        \nonumber
    \end{align}
    which can be shown using the relationship $\|\Psi_2\circ \Psi_1-\Phi_2\circ \Phi_1 \|_{\diamond}\leq \|\Psi_2-\Phi_2\|_{\diamond}+\|\Psi_1-\Phi_1\|_{\diamond}$ for arbitrary CPTP maps $\Psi_1, \Phi_1\in [\mathcal{L}(\mathcal{H}_0)\to \mathcal{L}(\mathcal{H}_1)]$ and $\Psi_2, \Phi_2\in [\mathcal{L}(\mathcal{H}_1)\to \mathcal{L}(\mathcal{H}_2)]$ where $\mathcal{H}_0$, $\mathcal{H}_1$, $\mathcal{H}_2$ are Hilbert spaces \cite{watrous2018theory}. 
    Also, $\mathcal{E}^{(3)}\circ \mathcal{F}\circ \mathcal{E}^{(1)}$ satisfies
    \begin{align}
        (\mathcal{E}^{(3)}\circ \mathcal{F}\circ \mathcal{E}^{(1)})(\rho)&=
        (\mathcal{E}^{(3)}\circ \mathcal{F})(\ket{0}\bra{0}\otimes \rho)
        \nonumber\\
        &=
        \mathcal{E}^{(3)}(\ket{0}\bra{0}\otimes e^{-if(H)t}\rho e^{if(H)t})
        \nonumber\\
        &=
        e^{-if(H)t}\rho e^{if(H)t} ,
        \nonumber
    \end{align}
    which completes the proof.
    
    \item [Variance]
    Denoting the quantum operation performed in steps 3-7 of Algorithm 1 when $\vec{j}$ is chosen as $\mathcal{E}^{(2)}_{\vec{j}}$, we obtain
    \begin{align}
        \underset{\substack{ \mathrm{dim}(\mathcal{H}^{\prime}) \\ \ket{\psi}\in \mathcal{H}\otimes \mathcal{H}^{\prime}; \\ \|\ket{\psi}\|=1 }}{\mathrm{sup}}
            \sum_{\vec{j}} p_{\vec{j}} \|
            &((\mathcal{E}^{(3)}\circ \mathcal{F}\circ \mathcal{E}^{(1)}-
            \mathcal{E}^{(3)}\circ \mathcal{E}^{(2)}_{\vec{j}}\circ
            \mathcal{E}^{(1)} )
            \nonumber\\
            &
            \otimes I_{\mathcal{H}^{\prime}})
            (\ket{\psi}\bra{\psi})
            \|_1^2
            \nonumber\\
            =
            \underset{\substack{ \mathrm{dim}(\mathcal{H}^{\prime}) \\ \ket{\psi}\in \mathcal{H}\otimes \mathcal{H}^{\prime}; \\ \|\ket{\psi}\|=1 }}{\mathrm{sup}}
            \sum_{\vec{j}} p_{\vec{j}} \|
            &((\mathcal{E}^{(3)}\circ \mathcal{F} -
            \mathcal{E}^{(3)}\circ \mathcal{E}^{(2)}_{\vec{j}}
            )\otimes I_{\mathcal{H}^{\prime}})
            \nonumber\\
            &(\ket{0}\bra{0}_{\mathcal{H}_c}\otimes\ket{\psi}\bra{\psi})
            \|_1^2
            \nonumber\\
            \leq
            \underset{\substack{ \mathrm{dim}(\mathcal{H}^{\prime}) \\ \ket{\phi}\in \mathcal{H}_c\otimes \mathcal{H}\otimes \mathcal{H}^{\prime}; \\ \|\ket{\phi}\|=1 }}{\mathrm{sup}}
            \sum_{\vec{j}} p_{\vec{j}} \|
            &((\mathcal{E}^{(3)}\circ \mathcal{F} -
            \mathcal{E}^{(3)}\circ \mathcal{E}^{(2)}_{\vec{j}}
            )\otimes I_{\mathcal{H}^{\prime}})
            \nonumber\\
            &(\ket{\phi}\bra{\phi})
            \|_1^2
            \nonumber\\
            =
            \underset{\substack{ \mathrm{dim}(\mathcal{H}^{\prime}) \\ \ket{\phi}\in \mathcal{H}_c\otimes \mathcal{H}\otimes \mathcal{H}^{\prime}; \\ \|\ket{\phi}\|=1 }}{\mathrm{sup}}
            \sum_{\vec{j}} p_{\vec{j}} \|
            &\mathrm{tr}_{\mathcal{H}_c}[
            ((\mathcal{F} - \mathcal{E}^{(2)}_{\vec{j}})\otimes I_{\mathcal{H}^{\prime}})(\ket{\phi}\bra{\phi})]
            \|_1^2
            \nonumber\\
            \leq
            \underset{\substack{ \mathrm{dim}(\mathcal{H}^{\prime}) \\ \ket{\phi}\in \mathcal{H}_c\otimes \mathcal{H}\otimes \mathcal{H}^{\prime}; \\ \|\ket{\phi}\|=1 }}{\mathrm{sup}}
            \sum_{\vec{j}} p_{\vec{j}} \|
            &
            ((\mathcal{F} - \mathcal{E}^{(2)}_{\vec{j}})\otimes I_{\mathcal{H}^{\prime}})(\ket{\phi}\bra{\phi})
            \|_1^2\leq 4\epsilon .
            \label{third_inequality}
    \end{align}
    The second inequality in Eq.\ (\ref{third_inequality}) is shown by the fact that the partial trace is a 1-norm non-increasing map, which can be proved by
    \begin{align}
        \|\mathrm{tr}_{\mathcal{H}_B}(M)\|_1
        &=\underset{U_A\in \mathcal{L}(\mathcal{H}_A)}{\mathrm{sup}}|\mathrm{tr}_{\mathcal{H}_A}(\mathrm{tr}_{\mathcal{H}_B}(M)U_A)|
        \nonumber\\
        =&\underset{U_A\in \mathcal{L}(\mathcal{H}_A)}{\mathrm{sup}}
        |\mathrm{tr}_{\mathcal{H}_A\otimes \mathcal{H}_B}(M(U_A\otimes I_B))|
        \nonumber\\
        \leq &\underset{U_{AB}\in \mathcal{L}(\mathcal{H}_A\otimes\mathcal{H}_B)}{\mathrm{sup}}|\mathrm{tr}_{\mathcal{H}_A\otimes \mathcal{H}_B}(MU_{AB})|
        =\|M\|_1 ,
        \nonumber
    \end{align}
    where $\mathcal{H}_A, \mathcal{H}_B$ are Hilbert spaces, $M$ is an operator in $\mathcal{L}(\mathcal{H}_A\otimes \mathcal{H}_B)$, and $U_A, U_{AB}$ are unitaries. The third inequality in Eq.\ (\ref{third_inequality}) follows from Eq.\ (\ref{variance_def}) because $\mathcal{F}$ is a Class T transformation.
\end{description}
\qed

\section{Appendix B: General relationship between error and variance}

In this appendix, we provide a theorem on the relationship between the error and variance of a general random protocol for approximating a unitary operation.

\begin{Theorem}\label{theo:2}
    For an arbitrary unitary operation defined by $\mathcal{U}(\rho ):=U\rho U^{\dagger}$ with a unitary $U$ and a density operator $\rho$ on a Hilbert space $\mathcal{H}$, if a set of deterministic quantum operations (completely-positive trace-preserving maps) $\mathcal{F}_j:\mathcal{L}(\mathcal{H})\to \mathcal{L}(\mathcal{H})$ and a probability distribution $\{ p_j \}$ satisfies
    \begin{align}
        \label{given}
        \| \mathcal{U}-\sum_j p_j \mathcal{F}_j \|_{\diamond}\leq \Delta
    \end{align}
    for some $\Delta >0$, the inequality  
    \begin{align}
        \label{simesi}
        \sum_j p_j\|\mathcal{U}(\ket{\psi}\bra{\psi})-
        \mathcal{F}_j(\ket{\psi}\bra{\psi}) \| _1 ^2
        \leq 2\Delta
    \end{align}
holds for any pure state $\ket{\psi}\in \mathcal{H}$.
\end{Theorem}
Note that the equation
\begin{align}
    \| \mathcal{U}\otimes \mathcal{I}-\sum_j p_j \mathcal{F}_j\otimes \mathcal{I} \|_{\diamond}\leq \Delta
    \nonumber
\end{align}
follows from the properties of the diamond norm \cite{watrous2018theory} given by Eq.\ (\ref{given}), thus Theorem 2 can be strengthened to 
\begin{Theorem}
    For an arbitrary unitary operation defined by $\mathcal{U}(\rho ):=U\rho U^{\dagger}$ with an unitary operator $U$ and a density operator $\rho$ on a Hilbert space $\mathcal{H}$, if a set of deterministic quantum operations (completely-positive trace-preserving maps) $\mathcal{F}_j:\mathcal{L}(\mathcal{H})\to \mathcal{L}(\mathcal{H})$ and a probability distribution $\{ p_j \}$ satisfies
    \begin{align}
        \| \mathcal{U}-\sum_j p_j \mathcal{F}_j \|_{\diamond}\leq \Delta
        \nonumber
    \end{align}
    for some $\Delta >0$, the variance of the averaged operation $\sum_j p_j \mathcal{F}_j$ is upper bounded by 
    \begin{align}
        \underset{\substack{ \mathrm{dim}(\mathcal{H}^{\prime}) \\ \ket{\psi};\|\ket{\psi}\|=1 }}{\mathrm{sup}}
        \sum_j p_j\|\mathcal{U}\otimes \mathcal{I}(\ket{\psi}\bra{\psi})-
        \mathcal{F}_j\otimes \mathcal{I}(\ket{\psi}\bra{\psi}) \| _1 ^2
        \leq 2\Delta
    \end{align}
    where $\mathcal{I}$ is the identity operation on a Hilbert space $\mathcal{H}^{\prime}$ 
    and $\ket{\psi}$ is a pure state on $\mathcal{H}\otimes \mathcal{H}^{\prime}$.
\end{Theorem}

As will be shown in the proof, Equation (\ref{simesi}) still holds when the condition for the bound of the error in Eq.~(\ref{given}) is weakened to only hold for pure states $\ket{\psi}\in \mathcal{H}$ as
\begin{align}
    \label{weak}
    \| \mathcal{U}(\ket{\psi}\bra{\psi})-\sum_j p_j \mathcal{F}_j(\ket{\psi}\bra{\psi}) \|_1 \leq \Delta,
\end{align}
instead of the case of the diamond norm as in Eq.~(\ref{given}).

First, we rewrite the statement of Theorem \ref{theo:2} by introducing a set of deterministic quantum operations $\mathcal{G}_j$ as
\begin{align}
    \mathcal{G}_j := \mathcal{U}^{-1}\circ \mathcal{F}_j.
    \nonumber
\end{align}
The following Lemma for the quantum operation $\sum_j p_j \mathcal{G}_j$ implemented by a random protocol applying $\mathcal{G}_j$ with probability $p_j$ provides a sufficient condition for Theorem \ref{theo:2}. 

\begin{lem}\label{lem:reduced}
If the error of a random protocol $\sum_j p_j \mathcal{G}_j $ simulating an identity operation on any pure state $\ket{\psi} \in \mathcal{H}$ satisfies
\begin{align}
    \label{joken}
    \| \ket{\psi}\bra{\psi} - \sum_j p_j \mathcal{G}_j(\ket{\psi}\bra{\psi}) \|_{1} \leq \Delta,
\end{align}
then the bound of the variance satisfies
\begin{align}
    \label{kaisimesi}
    \sum_j p_j\|\ket{\psi}\bra{\psi}-
    \mathcal{G}_j(\ket{\psi}\bra{\psi})\| _1 ^2
    \leq 2\Delta.
\end{align}
\end{lem}

The sufficiency of this statement is shown at the end of this appendix. We also define a complex coefficient $a_j := \bra{\psi}\mathcal{G}_j (\ket{\psi}\bra{\psi})\ket{\psi}$ to decompose the action of $\mathcal{G}_j$ on any pure state $\ket{\psi}$ as
\begin{eqnarray}
\label{decomposition}
    \mathcal{G}_j (\ket{\psi}\bra{\psi})=
    a_j\ket{\psi}\bra{\psi}+(1-a_j)\rho _j
\end{eqnarray}
where 
$\rho_j$ is an operator satisfying $\mathrm{tr}(\rho _j)=1$ that is
perpendicular to $\ket{\psi}$, namely, $\bra{\psi}\rho _j\ket{\psi}=0$.

According to the Fuchs-van de Graaf inequalities\cite{fuchs1999cryptographic}, we have
\begin{align}
    \label{fuchs}
    \| \rho -\ket{\psi}\bra{\psi}\|_1
    \leq 2\sqrt{1-a}
\end{align}
for a density operator $\rho\in \mathcal{L}(\mathcal{H})$, a pure state $\ket{\psi}\in\mathcal{H}$, and $a:=\bra{\psi}\rho \ket{\psi}$.

Now, we are ready to prove Lemma \ref{lem:reduced}:
        
\textbf{Proof of Lemma \ref{lem:reduced}:}
First, we find a lower bound to the left hand side of Eq. (\ref{joken}). 
Using the decomposition of $\mathcal{G}_j (\ket{\psi}\bra{\psi})$ given by Eq.~(\ref{decomposition}), Eq. (\ref{joken}) can be bounded as 
\begin{eqnarray}
    \Delta &\geq &
    \| \ket{\psi}\bra{\psi}-\sum_j p_j \mathcal{G}_j(\ket{\psi}\bra{\psi}) \|_{1}
    \nonumber\\
    &=&
    \| \sum_j p_j((1-a_j)\ket{\psi}\bra{\psi}-
    (1-a_j)\rho_j
    ) \|_{1}
    \nonumber\\
    &\geq &
    \bra{\psi}
    \left[
        \sum_j p_j((1-a_j)\ket{\psi}\bra{\psi}-
    (1-a_j)\rho_j
    \right]
    \ket{\psi}\nonumber\\
    &+&\left|\sum_k \bra{\psi_{k}^{\perp}}
    \left[
        \sum_j p_j((1-a_j)\ket{\psi}\bra{\psi}-
    (1-a_j)\rho_j
    \right]
    \ket{\psi_{k}^{\perp}}\right |
        \nonumber\\
    \label{geqgeq}
    &=&(1-\sum_jp_ja_j)+(1-\sum_jp_ja_j)=2(1-\sum_jp_ja_j) \nonumber \\
\end{eqnarray}
where {$\{ \ket{\psi_j^\perp} \}$} is a set of states (density operators) perpendicular to $\ket{\psi}$ and the set {$\{ \ket{\psi},  \ \ket{\psi_1^\perp}, \ket{\psi_2^\perp} \cdots \}$} forms an orthonormal basis. 
In the above evaluation, the second inequality holds due to the definition $$\|A \|_1 =\underset{U:\text{unitary}}{\mathrm{sup}}|\mathrm{tr}(AU)|,$$ namely, this is based on the fact that for a unitary
\begin{equation}
U=\ket{\psi}\bra{\psi}-\sum_k \mathrm{sgn}(\bra{\psi_k^{\perp}}\sum_jp_j(1-a_j)\rho_j\ket{\psi_k^{\perp}})\ket{\psi_k^{\perp}}\bra{\psi_k^{\perp}},  
\end{equation}
the term
\begin{equation}
	|\mathrm{tr}([\sum_j p_j((1-a_j)\ket{\psi}\bra{\psi}-(1-a_j)\rho_j)]U)|	
\end{equation}
is equal to the right hand side of the second inequality.

Next, we find a upper bound of the left hand side of Eq. (\ref{kaisimesi}).
Using Eq. (\ref{fuchs}), the left hand side of Eq.~(\ref{kaisimesi}) can be bound as
\begin{align}
    \sum_j p_j\|\ket{\psi}\bra{\psi}-
    \mathcal{G}_j(\ket{\psi}\bra{\psi})\| _1 ^2
    &\leq 
    \sum_j p_j (2\sqrt{1-a_j})^2
    \nonumber\\
    \label{hutatsume}
    &=4(1-\sum_jp_ja_j).  
\end{align}
Combining Eq.~(\ref{geqgeq}) and Eq.~(\ref{hutatsume}), 
we obtain
\begin{align}
    \sum_j p_j\|\ket{\psi}\bra{\psi}-
    \mathcal{G}_j(\ket{\psi}\bra{\psi})\| _1 ^2
    &\leq 4(1-\sum_jp_ja_j)
    \nonumber\\
    &\leq 2\Delta,
    \nonumber
\end{align}
which proves Eq.~(\ref{kaisimesi}).\qed

\textbf{Proof of Theorem \ref{theo:2}.}
The sufficiency of the statement (Eq.~(\ref{joken}) $\Rightarrow$ Eq.~(\ref{kaisimesi})) for proving Theorem \ref{theo:2} is based on the equivalence of Eq.~(\ref{weak}) (the weaker version of Eq.~(\ref{given})) and Eq.~(\ref{joken}), and the equivalence of Eq.~(\ref{simesi}) and Eq.~(\ref{kaisimesi}). The one-norm $\|\cdot \|_1$ is invariant under unitary transformations, namely, for an arbitrary linear operator $A$ and a unitary operator $V^{\prime}$, $\|A\|_1=\|V^{\prime}A\|_1=\|AV^{\prime}\|_1$ holds. This is because $$\underset{V:\text{unitary}}{\mathrm{sup}}|\mathrm{tr}(AV)|=\underset{V:\text{unitary}}{\mathrm{sup}}|\mathrm{tr}(V^{\prime}AV)|=\underset{V:\text{unitary}}{\mathrm{sup}}|\mathrm{tr}(AV^{\prime}V)|$$ holds. Therefore, the left hand side of Eq.~(\ref{weak}) and the left hand side of Eq.~(\ref{joken}) can be shown to be equal as
\begin{align}
    &\| \mathcal{U}(\ket{\psi}\bra{\psi})-\sum_j p_j \mathcal{F}_j(\ket{\psi}\bra{\psi}) \|_1  \nonumber \\
    = 
    &\left\| U^{\dagger} \left( \mathcal{U}(\ket{\psi}\bra{\psi})-\sum_j p_j \mathcal{F}_j(\ket{\psi}\bra{\psi}) \right) U \right\|_1 
    \nonumber \\
    =
    &\left\|  \mathcal{U}^{-1} \circ \mathcal{U}(\ket{\psi}\bra{\psi})-\sum_j p_j \mathcal{U}^{-1} \circ \mathcal{F}_j(\ket{\psi}\bra{\psi}) \right\|_1 
    \nonumber \\
    =
    &\left\|  \ket{\psi}\bra{\psi}-\sum_j p_j \mathcal{G}_j(\ket{\psi}\bra{\psi}) \right\|_1 .
    \nonumber
\end{align}
The equality of the left-hand side of Eq.~(\ref{simesi}) and the left-hand side of Eq.~(\ref{kaisimesi}) can be shown in the same way. Thus the equivalence of Eq.~(\ref{weak}) and Eq.~(\ref{joken}), and the equivalence of Eq.~(\ref{simesi}) and Eq.~(\ref{kaisimesi}) can be shown.\qed

\section{Appendix C: Universality of Algorithm 1}
In this section, we show that our algorithm simulating $e^{-iH\tau}\mapsto e^{-if(H)t}$ $(\tau ,t>0 )$ is a universal algorithm to simulate physically realizable linear transformation on a Hamiltonian. Under the assumption that $f(H)$ can also be seen as a Hamiltonian if an input $H$ is a Hamiltonian, we can assume that $f$ is a hermitian-preserving linear map. The universality of our algorithm can be shown in the following lemma. 
\begin{lem}
    The following two classes of linear maps are equal:
    \begin{enumerate}
        \renewcommand{\labelenumi}{(\alph{enumi})}
        \item The class of hermitian-preserving linear maps $f:\mathcal{L}(\mathcal{H})\mapsto \mathcal{L}(\mathcal{H})$ such that the transformation $e^{-iH\tau}\mapsto e^{-if(H)t}$ $(\tau ,t>0 )$ is physically realizable with an arbitrarily small error
        \item The class of hermitian-preserving linear maps $f:\mathcal{L}(\mathcal{H})\mapsto \mathcal{L}(\mathcal{H})$ such that $f(I)\propto I$
    \end{enumerate}
\end{lem}
Since our algorithm can simulate an arbitrary $f$ in class (b), it is shown to be able to simulate arbitrary ``physically realizable'' $f$. 

\Proof
    The Algorithm 1 can simulate $e^{-iH\tau}\mapsto e^{-if(H)t}$ $(\tau ,t>0 )$ for $f$ such that $f(I)\propto I$, thus (b)$\subseteq$(a). Assuming that $e^{-iH\tau}\mapsto e^{-if(H)t}$ $(\tau ,t>0 )$ is physically realizable for an $f$ such that $f(I)$ is not proportional to $I$, the output unitaries $e^{-if(H_1)t}=I$, $e^{-f(H_2)t}$ for $H_1:=0, H_2:=I$ are physically distinguishable. However, the inputs $e^{-iH_1\tau}=I$ and $e^{-iH_2\tau}=e^{-i\tau}I$ are only different up to the global phase and thus physically indistinguishable, which leads to a contradiction. Thus (a)$\subseteq$(b) is proved.
    \qed

\section{Appendix D: The application of simulating the negative time-evolution to Hamiltonian block encoding}

One application of simulating the negative time-evolution is the block-encoding of an unknown operator given as a block of an unknown Hamiltonian.  In this appendix, we present an algorithm for Hamiltonian block encoding utilizing our algorithm and analyze the approximation errors of the algorithm. 

\subsubsection{Algorithm for Hamiltonian block encoding}

Assume we are given access to the Hamiltonian dynamics $e^{-iH(A)\tau}$ $(\tau >0)$ of a Hamiltonian $H(A)$ with an upper bound $\Delta_{H(A)}$ of the maximum difference in energy eigenvalues represented as
\begin{align}
	H(A):=
	\left(
	\begin{array}{cc}
		\cdot&A^{\dagger}\\
		A&\cdot \\
	\end{array}
	\right) ,
	\nonumber
\end{align}
where diagonal blocks can be of arbitrary operators, and the smallest singular value $\lambda_{\mathrm{min}}$ of the operator $A$ on the off-diagonal part is positive. Then, we can construct a quantum operation $\mathcal{E}$ approximating the operation of a unitary operator $U(A)$, giving a block-encoding of $A$ defined as
\begin{align}
	U(A):=
	i
	\left(
	\begin{array}{cc}
		\sqrt{I-A^{\dagger}A}&A^{\dagger}\\
		A&-\sqrt{I-AA^{\dagger}}\\
	\end{array}
	\right)
	\nonumber
\end{align}
in a runtime of $O({\beta^2 \Delta_{H(A)}^2(\log (1/\epsilon) )^2n}/{\epsilon \lambda_{\mathrm{min}}^{2}})$ with an allowed error
\begin{align}\label{error_op1}
	\underset{\rho}{\mathrm{max}}
	\|
	U(A)\rho U(A)^{\dagger} - \mathcal{E}(\rho )
	\|_{\mathrm{op}}\leq \epsilon \,.
\end{align}
If we know that $H(A)$ belongs to a subspace of $\mathcal{L}(\mathcal{H})$ spanned by $J\subset {\{0,1,2,3\}^n}$, then $\beta := 2|J|$.

This construction is realized by combining the algorithm presented in \cite{lloyd2021hamiltonian} with our algorithm simulating the negative time-evolution. The algorithm presented in \cite{lloyd2021hamiltonian} requires the use of $e^{-iH(A)\tau}$ for both positive and negative $\tau$, but $e^{-iH(A)\tau}$ for only positive $\tau$ is used in our algorithm. Therefore our algorithm broadens the applicability of the quantum singular value transformation \cite{gilyen2019quantum, martyn2021grand} to the case where the classical description of the target operator $A$ is unknown, but is given as the dynamics of a Hamiltonian whose off-diagonal block is guaranteed to be given by $A$. 

\subsubsection{Analysis of the runtime}

Ref.\ \cite{lloyd2021hamiltonian} gives an algorithm which simulates $U(A)$ from dynamics $e^{\pm iH^{\prime}(A)\tau}$ $(\tau >0)$ where $H^{\prime}(A)$ is defined as
\begin{align}
    H^{\prime}(A):=
    \left(
	\begin{array}{cc}
		0&A^{\dagger}\\
		A&0\\
	\end{array}
	\right),
	\nonumber
\end{align}
where $A$ is a linear operator satisfying $A^{\dagger}A\leq I$, 
in an error smaller than $\epsilon >0$ using $O(\frac{\log(1/\epsilon)}{\lambda_{\mathrm{min}}})$ queries in total to $e^{\pm iH^{\prime}(A)\tau}$. Moreover, $\tau$ can be fixed to $\tau =1$. This is achieved by constructing a unitary
\begin{align}
&\bigoplus_j\nonumber\\
&(\ket{r_j}\ \ket{l_j})
    \left(
	\begin{array}{cc}
		P(\cos \lambda_j)&i\sin \lambda_j Q(\cos \lambda_j)\\
		i\sin \lambda_j Q^*(\cos \lambda_j)&P^*(\cos \lambda_j)\\
	\end{array}
	\right)
  \left(
 \begin{array}{c}
		\bra{r_j}\\
		\bra{l_j}\\
\end{array}
\right)
	\nonumber
\end{align}
by performing the quantum singular value transformation using
\begin{align}
&\bigoplus_j
(\ket{r_j}\ \ket{l_j})
    \left(
	\begin{array}{cc}
		\cos \lambda_j&i\sin \lambda_j\\
		i\sin \lambda_j&\cos \lambda_j\\
	\end{array}
	\right)
 \left(
 \begin{array}{c}
		\bra{r_j}\\
		\bra{l_j}\\
\end{array}
\right)
 \nonumber\\
 =& e^{iH^{\prime}(A)}
	\nonumber
\end{align}
and its dagger $e^{-iH^{\prime}(A)}$,
where $A=\sum_j \lambda_j \ket{l_j}\bra{r_j}$ $(\lambda_{\mathrm{min}} \leq \lambda_j \leq 1)$ is a singular value decomposition of $A$, and polynomial functions $P$, $Q$ are chosen in such a way that $\sin \lambda_j Q^*(\cos \lambda_j)=\sqrt{1-\cos^2 \lambda_j} Q^*(\cos \lambda_j)$ approximates $\lambda_j$ and $-iP(\cos \lambda_j)$ approximates $\sqrt{1-\lambda_j^2}$ both with an error smaller than or equal to $\epsilon$ for all $\lambda_{\mathrm{min}} \leq \lambda_j \leq 1$ (equivalently, $\sqrt{1-x^2}Q^*(x)$ approximates $\mathrm{arccos}(x)$ and $-iP(x)$ approximates $\sqrt{1-\mathrm{arccos}(x)^2}$ both with an error smaller than or equal to $\epsilon$ for all $\cos 1\leq x\leq \cos \lambda_{\mathrm{min}} = 1-O(\lambda_{\mathrm{min}} ^2)$). The measure of the approximation error is based on the error of approximating $\mathrm{arccos}$ and is different from the measure used in Eq.\ (\ref{error_op1}), but it can be easily shown that the error in terms of Eq.\ (\ref{error_op1}) is bounded by 
\begin{align}
\underset{\substack{a_j,b_j,c_j,d_j\in \mathbb{C}\\ |a_j|,|b_j|,|c_j|,|d_j|\leq 1}}{\mathrm{max}}
\left\|
\bigoplus_j
\left(
	\begin{array}{cc}
		\epsilon a_j&\epsilon b_j\\
		\epsilon c_j&\epsilon d_j\\
	\end{array}
	\right)
 \right\| _{\mathrm{op}}\leq 4\epsilon \,,
 \nonumber
\end{align}
thus the total number of queries to $e^{\pm iH^{\prime}(A)}$ for the case where the allowed error in terms of Eq.\ (\ref{error_op1}) is chosen as $\epsilon$ is also $O(\frac{\log(1/\epsilon)}{\lambda_{\mathrm{min}}})=:d(\epsilon)$.

Suppose that we approximate $U(A)$ using this algorithm for the case where the allowed error in terms of Eq.\ (\ref{error_op1}) is set as $\epsilon /2$ using $e^{-iH^{\prime}(A)}$ and a quantum operation $\mathcal{E}^{\prime}$ which approximates the quantum operation of $e^{iH^{\prime}(A)}$ using $e^{-iH^{\prime}(A)}$ instead of preparing $e^{iH^{\prime}(A)}$. For implementing $\mathcal{E}^{\prime}$, we choose an allowed error in terms of Eq. (\ref{our_error}) to be $\epsilon/4d(\epsilon /2)$.
In this situation, the overall procedure only requires $e^{-iH^{\prime}(A)\tau}$ as input. Because the error measured by Eq. (\ref{our_error}) times two is larger than the error measured by Eq. (\ref{error_op1}) based on the fact the the diamond norm $\|P\|_{\diamond}$ of an operation $P$ is greater than or equal to the operator norm $\|P(\rho )\|_{\mathrm{op}}$ for an arbitrary input density operator $\rho$, the overall error of simulating $U(A)$ in terms of Eq. (\ref{error_op1}) is upper bounded by
\begin{align}
	\frac{\epsilon }{2}+
	2\cdot \frac{\epsilon }{4d(\epsilon /2)}\cdot d(\epsilon /2)=\epsilon ,
	\nonumber
\end{align}
where the second term of the left hand side corresponds to the total error arising from the approximation error for $\mathcal{E}^{\prime}$ calculated by (error of approximating $\mathcal{E}^{\prime}$) $\times$ (upper bound of the number of queries to $\mathcal{E}^{\prime}$). Because $\mathcal{E}^{\prime}$ can be constructed from $e^{-i\tau H^{\prime}(A)}$ by simulating $e^{iH^{\prime}(A)}$ with the negative time-evolution for $t=1$ with an allowed error in terms of  Eq. (\ref{our_error}) being chosen as $\epsilon/4d(\epsilon /2)$ in a runtime 
$O(\frac{\beta ^2 1^2 \Delta_{H'(A)}^2 n}{\epsilon/4d(\epsilon /2)})=O(\frac{\beta ^2 \Delta_{H'(A)}^2 nd(\epsilon /2)}{\epsilon})$ (see the main text), the total runtime of constructing $U(A)$ will be 

\begin{align}
O&\left(\frac{\beta ^2 \Delta_{H'(A)}^2  n d(\epsilon /2)}{\epsilon}\cdot d(\epsilon /2)\right)
=O\left(\frac{\beta ^2 \Delta_{H'(A)}^2 d(\epsilon /2)^2n}{\epsilon}\right) \nonumber \\
=&O\left(\frac{\beta ^2 \Delta_{H'(A)}^2 (\log (1/\epsilon))^2n}{\epsilon \lambda_{\mathrm{min}}^2}\right)
\end{align}

calculated by (runtime of simulating $\mathcal{E}^{\prime}$) $\times$ (upper bound of the number of queries to $\mathcal{E}^{\prime}$). 

We have described the runtime analysis of constructing $U(A)$ using $e^{-iH^{\prime}(A)\tau}$ so far. This procedure can be extended to the case where the input dynamics is $e^{-iH(A)\tau}$ instead of $e^{-iH^{\prime}(A)\tau}$ because $e^{\pm i H^{\prime}(A)t}$ can be constructed from $e^{-iH(A)\tau}$ and $e^{iH(A)\tau}$, and $e^{iH(A)\tau}$ is further constructed from $e^{-iH(A)\tau}$ by negative time-evolution based on the equation
\begin{align}
\pm \frac{1}{2}[H(A)+(Z\otimes I)(-H(A))(Z\otimes I)]=\pm H^{\prime}(A) .
\end{align}
Since $H(A)\mapsto -H(A)$ is simulated by a Class T transformation (see Appendix A), the above equation also provides the description of Class T transformations which transform $H(A)$ to $\pm H^{\prime}(A)$. Denoting this transformation as $H(A)\mapsto \sum_j h_j U_j H(A) U_j^{\dagger}$, the sum $\sum_j h_j$ is calculated as $(\beta + 1)/2$ which is $O(\beta)$. This technique is also introduced in \cite{lloyd2021hamiltonian}. Using this technique, quantum operations approximating $e^{\pm iH^{\prime}(A)}$ with an allowed error $\epsilon /4d(\epsilon /2)$ can be constructed from $e^{-iH(A)\tau}$ $(\tau >0)$ also in time $O(\frac{\beta ^2 \Delta_{H(A)}^2 nd(\epsilon /2)}{\epsilon})$, thus $U(A)$ can be constructed in runtime $O(\frac{\beta ^2\Delta_{H(A)}^2 (\log (1/\epsilon))^2 n}{\epsilon \lambda_{\mathrm{min}}^2})$ as well.

\section{Appendix E: Runtime and total evolution time analysis of Hamiltonian single-parameter learning}

In this appendix, we present the runtime and total evolution time analysis of our algorithm for the Hamiltonian single-parameter learning task of efficiently estimating a single parameter from the dynamics of a $n$-qubit unknown Hamiltonian $H=\sum_{\vec{u}} c_{\vec{u}} \sigma_{\vec{u}}$ satisfying $(|c_{\vec{u}}|\leq 1)$ with the Pauli vectors $(\vec{v}\in \{0,1,2,3\}^n)$ presented in the main text. 

In the Hamiltonian transformation algorithm (Algorithm 1) applied for the Hamiltonian single-parameter learning task, we choose the transformation of the Hamiltonian given by 
\begin{align}
f_{\vec{v}}(H)=c_{\vec{v}}Y\otimes I\otimes \cdots \otimes I.
\label{extract}
\end{align}
The set of linear operators $f_{\vec{v}}$ $(\vec{v}\in \{0,1,2,3\}^n)$ is 
characterized by $\gamma_{\vec{w},\vec{u}}$ defined as
\begin{align}
\gamma_{\vec{w},\vec{u}}:=
\delta_{\vec{w},(2,0,\ldots ,0)} \delta_{\vec{u},\vec{v}}.
\nonumber
\end{align}

\subsubsection{Probability distribution obtained by Algorithm 1}

We consider performing a projective measurement on the basis of $\{\ket{0},\ket{1}\}$ or $\{\ket{+},\ket{-}\}$ on the first qubit, namely, the qubit on which the $Y$ operator appears in Eq.\ (\ref{extract}) of the state $e^{-if_{\vec{v}}(H)t}(\ket{0}\otimes \ket{0}^{\otimes n-1})$. The probability of obtaining the outcome $0$ for the projective measurement in the basis of $\{\ket{0},\ket{1}\}$ is given by
\begin{align}
p_0:=
\mathrm{tr}[
&(\ket{0}\bra{0}\otimes I^{\otimes n-1})\cdot \nonumber\\
&\{
e^{-if_{\vec{v}}(H)t}(\ket{0}\bra{0}\otimes (\ket{0}\bra{0})^{\otimes n-1})
e^{if_{\vec{v}}(H)t}
\}
]\nonumber\\
=&
\mathrm{tr}[\ket{0}\bra{0}(e^{-ic_{\vec{v}}tY}\ket{0}\bra{0}(e^{ic_{\vec{v}}tY})]
\nonumber\\
=&\frac{1+\cos{(2c_{\vec{v}}t)}}{2}
\label{prob0}
\end{align}
and the probability of obtaining the outcome $+$ for the projective measurement in the basis of $\{\ket{+},\ket{-}\}$ is given by
\begin{align}
p_+:=
\mathrm{tr}[
&(\ket{+}\bra{+}\otimes I^{\otimes n-1})\cdot \nonumber\\
&\{
e^{-if_{\vec{v}}(H)t}(\ket{0}\bra{0}\otimes (\ket{0}\bra{0})^{\otimes n-1})
e^{if_{\vec{v}}(H)t}
\}
]\nonumber\\
=&
\mathrm{tr}[\ket{+}\bra{+}(e^{-ic_{\vec{v}}tY}\ket{0}\bra{0}(e^{ic_{\vec{v}}tY})]
\nonumber\\
=&\frac{1+\sin{(2c_{\vec{v}}t)}}{2}.
\label{prob+}
\end{align}

Suppose that we construct a quantum operation ${ \mathcal{F}}_{\mathrm{approx}}$ that approximates $\mathcal{F}(\rho):=e^{-if_{\vec{v}}(H)t}\rho e^{if_{\vec{v}}(H)t}$ with an allowed error $\epsilon /2$ in terms of Eq.\ (\ref{our_error}).

The runtime for simulating ${\mathcal{F}}_{\mathrm{approx}}$ is $O(\beta^2 t^2\Delta_H^2 n/\epsilon)$ for $\beta = 2$, while the total evolution time of $e^{-iH\tau}$ will only be $2t$, as can be seen from the procedure of Algorithm 1.

In this case, the probability of obtaining the outcome $0$ in the $\{\ket{0},\ket{1}\}$ basis measurement and $+$ in the $\{\ket{+},\ket{-}\}$ basis measurement will be close by $\epsilon$ to values in Eq.\ (\ref{prob0}) and Eq.\ (\ref{prob+}), respectively. 
This can be proved for the $+$ case (and similarly for the $0$ case) as
\begin{align}
&\left|\mathrm{tr}\left(\left[
\ket{+}\bra{+}\otimes I^{\otimes n-1}
\right]
{\mathcal{F}}_{\mathrm{approx}}\left(\ket{0}\bra{0}\otimes (\ket{0}\bra{0})^{\otimes n-1}\right)
\right)
\right.
\nonumber\\
&\left.
-\mathrm{tr}\left(\left[
\ket{+}\bra{+}\otimes I^{\otimes n-1}
\right]
\mathcal{F}\left(\ket{0}\bra{0}\otimes (\ket{0}\bra{0})^{\otimes n-1}\right)
\right)\right|\nonumber\\
=&
\left|
\sum_{j}\bra{+}\bra{j}\left[{ \mathcal{F}}_{\mathrm{approx}}\left(\ket{0}\bra{0}\otimes (\ket{0}\bra{0})^{\otimes n-1}\right)\right.\right.
\nonumber\\
&-
\left.\left.
\mathcal{F} \left(\ket{0}\bra{0}\otimes (\ket{0}\bra{0})^{\otimes n-1}\right)\right]
\ket{+}\ket{j}
\right|
\nonumber\\
\leq &
\sum_{a\in \{+, -\}, j}\left|\bra{a}\bra{j}\left[{ \mathcal{F}}_{\mathrm{approx}}\left(\ket{0}\bra{0}\otimes (\ket{0}\bra{0})^{\otimes n-1}\right)\right.\right.
\nonumber\\
&-
\left.\left.
\mathcal{F}\left(\ket{0}\bra{0}\otimes (\ket{0}\bra{0})^{\otimes n-1}\right)\right]
\ket{a}\ket{j}
\right|
\nonumber\\
=&\underset{s_{a,j}\in \{1, -1\}}{\mathrm{sup}}
\mathrm{tr}\left(
\left\{\sum_{a\in \{+, -\},j}s_{a,j}\ket{a}\bra{a}\otimes \ket{j}\bra{j}\right\}
\right.
\nonumber\\
&\quad \quad \quad \quad \quad \left.\left[(F_{\mathrm{approx}}- \mathcal{F})\left(\ket{0}\bra{0}\otimes (\ket{0}\bra{0})^{\otimes n-1}\right)\right]\right)
\nonumber\\
\leq &\underset{U:\text{unitary}}{\mathrm{sup}}
\mathrm{tr}\left(U
\left[({\mathcal{F}}_{\mathrm{approx}}- \mathcal{F})\left(\ket{0}\bra{0}\otimes (\ket{0}\bra{0})^{\otimes n-1}\right)\right]
\right)
\nonumber\\
=&
\left\|
({\mathcal{F}}_{\mathrm{approx}}- \mathcal{F})\left(\ket{0}\bra{0}\otimes (\ket{0}\bra{0})^{\otimes n-1}\right)
\right\|_1\leq \epsilon .
\nonumber
\end{align}
In the above equations, $\{\ket{j}\}_j$ is an orthonormal basis of the Hilbert space to which $I^{\otimes n-1}$ belongs.

\subsubsection{Learning $c_{\vec{v}}$ using robust quantum phase estimation }
Theorem I.1.\ in \cite{kimmel2015robust} can be rephrased to the following theorem.

\begin{Theorem}
    Suppose that we can perform two families of measurements, $(0,k)$-measurements and $(+,k)$-measurements, where $k\in \mathbb{Z}_{>0}$, whose success probabilities obtaining the outcome $0$ and outcome $+$ are given in terms of $\theta \in (-\pi, \pi]$ as
    \begin{align}
    \textbf{$(0,k)$-meas.:  }
    p_{0,k}(\theta)&:=\frac{1+\cos{(k\theta})}{2}+\delta_0(k),\nonumber\\
    \textbf{$(+,k)$-meas.:  }
    p_{+,k}(\theta)&:=\frac{1+\sin{(k\theta})}{2}+\delta_+(k) ,
    \nonumber
    \end{align}
    respectively, where $\delta_0(k)$ and $\delta_+(k)$ satisfy
    \begin{align}
    \underset{k}{\mathrm{sup}}\{
    |\delta_0(k)|, |\delta_+(k)|
    \} =:\delta_{\mathrm{sup}}<1/\sqrt{8} .
    \nonumber
    \end{align}

Then for any allowed standard deviation $s>0$, an estimate $\hat{ \theta}$ of $\theta$ can be obtained with a standard deviation smaller than or equal to $s$ by a classical computation with runtime $O(\mathrm{poly}K)$ of a function of the numbers of success of $(0,2^{j-1})$-measurements and $(+,2^{j-1})$-measurements $(j\in \{1,\ldots ,K\})$ both among $M_j$ times of the measurements. Here, $M_j$ and $K$ are defined as
    \begin{align}
    K&:=\mathrm{ceil}\left[\log_2{\left(\frac{3\pi}{s}\right)}\right]\label{KKK}\\
    M_j&:=F(\delta_{\mathrm{sup}})(3(K-j)+1)\nonumber\\
F(\delta_{\mathrm{sup}})&:=\mathrm{ceil}\left[
    \frac{\log \left( \frac{1}{2}(1-\sqrt{8}\delta_{\mathrm{sup}})\right)}{\log \left( 1-\frac{1}{2}(1-\sqrt{8}\delta_{\mathrm{sup}})^2\right)}
    \right] .\nonumber
    \end{align}
\end{Theorem}
We can perform $(0,2^{j-1})$-measurements and $(+,2^{j-1})$-measurements for $j\in \mathbb{Z}_{>0}$ for $\theta=c_{\vec{v}}$ using 
simulation of $e^{-if_{\vec{v}}(H)t}$ for $t=2^{j-2}$ and $\epsilon=1/2\sqrt{8}$  (or any positive number smaller than $1/\sqrt{8}$ can be used instead).
The runtime of $(0,2^{j-1})$-measurements and $(+,2^{j-1})$-measurements in this case is bounded above using a constant $C>0$ independent of $j$ by $C\Delta_H^2 4^{j-2}n$, and the total evolution time of the dynamics $e^{-iH\tau}$ is $2\cdot 2^{j-2}=2^{j-1}$. Therefore, the total time of running the quantum circuit is bounded above as

\begin{align}
\sum_{j=1}^{K} M_j\cdot C\Delta_H^2 4^{j-2}n=O(\Delta_H^24^Kn)=O\left(\frac{\Delta_H^2n}{s^2}\right) ,
\nonumber
\end{align}

while the total evolution time of $e^{-iH\tau}$ in the overall experiment is 
\begin{align}
\sum_{j=1}^{K} M_j2^{j-1}=O\left(\frac{1}{s}\right) .
\nonumber
\end{align}
In the above calculation, we use an equation
\begin{align}
\sum_{j=1}^{K}(K-j)r^{j}=\frac{r^{K+1}-Kr^2+(K-1)r}{(r-1)^2}
\nonumber
\end{align}
which holds for an arbitrary $r>0$.  

The runtime $O(\mathrm{poly}K)$ of the classical calculation can be ignored in the evaluation of the total runtime of estimating $c_{\vec{v}}$, since it depends on $s$ as $\mathrm{poly}\log (1/s)$ and tends to infinity more slowly than $O(1/s^2)$.

To calculate total evolution time in terms of the error upper bound $\epsilon >0$ and the maximum failure probability $\delta >0$, we can modify the learning procedure to repeating estimation of $c_{\vec{v}}$ with standard deviation smaller than or equal to $\epsilon /2$ for $O(\log (1/\delta))$ times and adopting the median of $O(\log (1/\delta))$ estimates. In this case, the total evolution time will be $O(\log (1/\delta)/\epsilon)$. The failure probability is shown to be smaller than or equal to $\delta$ using the fact that the probability that an estimate $\hat{c}_{\vec{v}}$ of $c_{\vec{v}}$ with standard deviation $\epsilon/2$ satisfies $|\hat{c}_{\vec{v}}-c_{\vec{v}}|>\epsilon$ is strictly smaller than $1/2$.

\subsubsection{Comparison with the Hamiltonian simulation method in \cite{huang2023learning}}

Recently, another Hamiltonian learning technique achieving the Heisenberg limit for precision scaling in parameter estimation of a low-interaction Hamiltonian has been proposed \cite{huang2023learning}. Given an $n$-qubit Hamiltonian $H=\sum_{\vec{v}\in J}c_{\vec{v}}\sigma_{\vec{v}}$ for a known set of vectors $J$ satisfying the low-interaction Hamiltonian condition, this technique estimates {\it all} values of $c_{\vec{v}}$ in a single run with the total evolution time $O((\log \delta )/\epsilon)$ where $\epsilon$ is precision and $\delta$ is the failure probability. 

Our method can learn {\it any} single parameter $c_{\vec{v}}$ of a {\it general} Hamiltonian for $\vec{v}\in \{0,1,2,3\}^n\backslash \{\vec{0}\}$ with total evolution time $O((\log \delta )/\epsilon)$, which also achieves the Heisenberg-limited precision scaling. 
Note that the total evolution time has no dependence on $\|H\|_{\rm op}$ or more generally on $n$. 
In this sense, we give a partial answer to the open problem proposed by \cite{huang2023learning} regarding the learning of Hamiltonians without any structure. We still note that in terms of the total {\it runtime} instead of the evolution time, our algorithm is only efficient when $\Delta_H$ is small.

For a low-interaction Hamiltonian, the estimation of all $c_{\vec{v}}$'s using our algorithm requires to run our algorithm for every parameter.   Therefore, the total evolution time is (${\rm poly}\ n$)$\times O((\log \delta )/\epsilon )$, which is longer than that of the algorithm in \cite{huang2023learning} by a polynomial factor (the number of parameters of the $n$-qubit low-interaction Hamiltonian). However, for estimation of a single parameter $c_{\vec{v}}$ for $\vec{v}$ representing the high-interaction Hamiltonian, namely, the term consisting of multiplications of $k=O(n)$ non-identity Pauli operators, the algorithm in \cite{huang2023learning} requires a total evolution time exponential in $k$, which leads to an exponential overall runtime, as can be seen from section C.2 of \cite{huang2023learning} (the algorithm in \cite{huang2023learning} can be applied to a general $n$-qubit Hamiltonian and restricted to estimation of only a single parameter). On the other hand, our algorithm requires a constant total evolution time $O((\log \delta )/\epsilon)$, and a runtime independent of $k$. 
For these reasons, our algorithm is suited to estimate a single parameter of non-local Hamiltonians. With our method, a parameter of an experimentally simulated Hamiltonian with a not-too-large norm, which does not necessarily have a simple structure, becomes obtainable.

Finally, we note that our algorithm also runs in a shorter runtime than the methods based on unitary tomography for the estimation of a single parameter. Recently, a unitary tomography method for estimating only a small number of entries of a unitary operation in a short time has been proposed \cite{haah2023query}.  However, this is not equivalent to the estimation of a small number of entries of a
Hamiltonian. In order to obtain the value of $c_{\vec{v}}$ of a Hamiltonian $H$ by tomography of the unitary evolution $e^{-iHt}$, the full-tomography of $e^{-iHt}$ is required, which requires a runtime in $O(4^n/\epsilon)$ \cite{haah2023query}.

\clearpage


\begin{thebibliography}{43}%
\makeatletter
\providecommand \@ifxundefined [1]{%
 \@ifx{#1\undefined}
}%
\providecommand \@ifnum [1]{%
 \ifnum #1\expandafter \@firstoftwo
 \else \expandafter \@secondoftwo
 \fi
}%
\providecommand \@ifx [1]{%
 \ifx #1\expandafter \@firstoftwo
 \else \expandafter \@secondoftwo
 \fi
}%
\providecommand \natexlab [1]{#1}%
\providecommand \enquote  [1]{``#1''}%
\providecommand \bibnamefont  [1]{#1}%
\providecommand \bibfnamefont [1]{#1}%
\providecommand \citenamefont [1]{#1}%
\providecommand \href@noop [0]{\@secondoftwo}%
\providecommand \href [0]{\begingroup \@sanitize@url \@href}%
\providecommand \@href[1]{\@@startlink{#1}\@@href}%
\providecommand \@@href[1]{\endgroup#1\@@endlink}%
\providecommand \@sanitize@url [0]{\catcode `\\12\catcode `\$12\catcode
  `\&12\catcode `\#12\catcode `\^12\catcode `\_12\catcode `\%12\relax}%
\providecommand \@@startlink[1]{}%
\providecommand \@@endlink[0]{}%
\providecommand \url  [0]{\begingroup\@sanitize@url \@url }%
\providecommand \@url [1]{\endgroup\@href {#1}{\urlprefix }}%
\providecommand \urlprefix  [0]{URL }%
\providecommand \Eprint [0]{\href }%
\providecommand \doibase [0]{https://doi.org/}%
\providecommand \selectlanguage [0]{\@gobble}%
\providecommand \bibinfo  [0]{\@secondoftwo}%
\providecommand \bibfield  [0]{\@secondoftwo}%
\providecommand \translation [1]{[#1]}%
\providecommand \BibitemOpen [0]{}%
\providecommand \bibitemStop [0]{}%
\providecommand \bibitemNoStop [0]{.\EOS\space}%
\providecommand \EOS [0]{\spacefactor3000\relax}%
\providecommand \BibitemShut  [1]{\csname bibitem#1\endcsname}%
\let\auto@bib@innerbib\@empty
%</preamble>
\bibitem [{\citenamefont {Suzuki}(1990)}]{suzuki1990fractal}%
  \BibitemOpen
  \bibfield  {author} {\bibinfo {author} {\bibfnamefont {M.}~\bibnamefont
  {Suzuki}},\ }\bibfield  {title} {\bibinfo {title} {Fractal decomposition of
  exponential operators with applications to many-body theories and {M}onte
  {C}arlo simulations},\ }\href@noop {} {\bibfield  {journal} {\bibinfo
  {journal} {Physics Letters A}\ }\textbf {\bibinfo {volume} {146}},\ \bibinfo
  {pages} {319} (\bibinfo {year} {1990})}\BibitemShut {NoStop}%
\bibitem [{\citenamefont {Suzuki}(1991)}]{suzuki1991general}%
  \BibitemOpen
  \bibfield  {author} {\bibinfo {author} {\bibfnamefont {M.}~\bibnamefont
  {Suzuki}},\ }\bibfield  {title} {\bibinfo {title} {General theory of fractal
  path integrals with applications to many-body theories and statistical
  physics},\ }\href@noop {} {\bibfield  {journal} {\bibinfo  {journal} {Journal
  of Mathematical Physics}\ }\textbf {\bibinfo {volume} {32}},\ \bibinfo
  {pages} {400} (\bibinfo {year} {1991})}\BibitemShut {NoStop}%
\bibitem [{\citenamefont {Campbell}(2019)}]{campbell2019random}%
  \BibitemOpen
  \bibfield  {author} {\bibinfo {author} {\bibfnamefont {E.}~\bibnamefont
  {Campbell}},\ }\bibfield  {title} {\bibinfo {title} {Random compiler for fast
  {H}amiltonian simulation},\ }\href@noop {} {\bibfield  {journal} {\bibinfo
  {journal} {Physical Review Letters}\ }\textbf {\bibinfo {volume} {123}},\
  \bibinfo {pages} {070503} (\bibinfo {year} {2019})}\BibitemShut {NoStop}%
\bibitem [{\citenamefont {Berry}\ \emph {et~al.}(2015)\citenamefont {Berry},
  \citenamefont {Childs}, \citenamefont {Cleve}, \citenamefont {Kothari},\ and\
  \citenamefont {Somma}}]{berry2015simulating}%
  \BibitemOpen
  \bibfield  {author} {\bibinfo {author} {\bibfnamefont {D.~W.}\ \bibnamefont
  {Berry}}, \bibinfo {author} {\bibfnamefont {A.~M.}\ \bibnamefont {Childs}},
  \bibinfo {author} {\bibfnamefont {R.}~\bibnamefont {Cleve}}, \bibinfo
  {author} {\bibfnamefont {R.}~\bibnamefont {Kothari}},\ and\ \bibinfo {author}
  {\bibfnamefont {R.~D.}\ \bibnamefont {Somma}},\ }\bibfield  {title} {\bibinfo
  {title} {Simulating {H}amiltonian dynamics with a truncated {T}aylor
  series},\ }\href@noop {} {\bibfield  {journal} {\bibinfo  {journal} {Physical
  Review Letters}\ }\textbf {\bibinfo {volume} {114}},\ \bibinfo {pages}
  {090502} (\bibinfo {year} {2015})}\BibitemShut {NoStop}%
\bibitem [{\citenamefont {Low}\ and\ \citenamefont
  {Chuang}(2017)}]{low2017optimal}%
  \BibitemOpen
  \bibfield  {author} {\bibinfo {author} {\bibfnamefont {G.~H.}\ \bibnamefont
  {Low}}\ and\ \bibinfo {author} {\bibfnamefont {I.~L.}\ \bibnamefont
  {Chuang}},\ }\bibfield  {title} {\bibinfo {title} {Optimal {H}amiltonian
  simulation by quantum signal processing},\ }\href@noop {} {\bibfield
  {journal} {\bibinfo  {journal} {Physical Review Letters}\ }\textbf {\bibinfo
  {volume} {118}},\ \bibinfo {pages} {010501} (\bibinfo {year}
  {2017})}\BibitemShut {NoStop}%
\bibitem [{\citenamefont {Low}\ and\ \citenamefont
  {Chuang}(2019)}]{low2019hamiltonian}%
  \BibitemOpen
  \bibfield  {author} {\bibinfo {author} {\bibfnamefont {G.~H.}\ \bibnamefont
  {Low}}\ and\ \bibinfo {author} {\bibfnamefont {I.~L.}\ \bibnamefont
  {Chuang}},\ }\bibfield  {title} {\bibinfo {title} {Hamiltonian simulation by
  qubitization},\ }\href@noop {} {\bibfield  {journal} {\bibinfo  {journal}
  {Quantum}\ }\textbf {\bibinfo {volume} {3}},\ \bibinfo {pages} {163}
  (\bibinfo {year} {2019})}\BibitemShut {NoStop}%
\bibitem [{\citenamefont {Childs}\ \emph {et~al.}(2018)\citenamefont {Childs},
  \citenamefont {Maslov}, \citenamefont {Nam}, \citenamefont {Ross},\ and\
  \citenamefont {Su}}]{childs2018toward}%
  \BibitemOpen
  \bibfield  {author} {\bibinfo {author} {\bibfnamefont {A.~M.}\ \bibnamefont
  {Childs}}, \bibinfo {author} {\bibfnamefont {D.}~\bibnamefont {Maslov}},
  \bibinfo {author} {\bibfnamefont {Y.}~\bibnamefont {Nam}}, \bibinfo {author}
  {\bibfnamefont {N.~J.}\ \bibnamefont {Ross}},\ and\ \bibinfo {author}
  {\bibfnamefont {Y.}~\bibnamefont {Su}},\ }\bibfield  {title} {\bibinfo
  {title} {Toward the first quantum simulation with quantum speedup},\
  }\href@noop {} {\bibfield  {journal} {\bibinfo  {journal} {Proceedings of the
  National Academy of Sciences}\ }\textbf {\bibinfo {volume} {115}},\ \bibinfo
  {pages} {9456} (\bibinfo {year} {2018})}\BibitemShut {NoStop}%
\bibitem [{\citenamefont {Lloyd}\ \emph {et~al.}(2021)\citenamefont {Lloyd},
  \citenamefont {Kiani}, \citenamefont {Arvidsson-Shukur}, \citenamefont
  {Bosch}, \citenamefont {De~Palma}, \citenamefont {Kaminsky}, \citenamefont
  {Liu},\ and\ \citenamefont {Marvian}}]{lloyd2021hamiltonian}%
  \BibitemOpen
  \bibfield  {author} {\bibinfo {author} {\bibfnamefont {S.}~\bibnamefont
  {Lloyd}}, \bibinfo {author} {\bibfnamefont {B.~T.}\ \bibnamefont {Kiani}},
  \bibinfo {author} {\bibfnamefont {D.~R.}\ \bibnamefont {Arvidsson-Shukur}},
  \bibinfo {author} {\bibfnamefont {S.}~\bibnamefont {Bosch}}, \bibinfo
  {author} {\bibfnamefont {G.}~\bibnamefont {De~Palma}}, \bibinfo {author}
  {\bibfnamefont {W.~M.}\ \bibnamefont {Kaminsky}}, \bibinfo {author}
  {\bibfnamefont {Z.-W.}\ \bibnamefont {Liu}},\ and\ \bibinfo {author}
  {\bibfnamefont {M.}~\bibnamefont {Marvian}},\ }\bibfield  {title} {\bibinfo
  {title} {Hamiltonian singular value transformation and inverse block
  encoding},\ }\href@noop {} {\bibfield  {journal} {\bibinfo  {journal} {arXiv
  preprint arXiv:2104.01410}\ } (\bibinfo {year} {2021})}\BibitemShut {NoStop}%
\bibitem [{\citenamefont {Granade}\ \emph {et~al.}(2012)\citenamefont
  {Granade}, \citenamefont {Ferrie}, \citenamefont {Wiebe},\ and\ \citenamefont
  {Cory}}]{granade2012robust}%
  \BibitemOpen
  \bibfield  {author} {\bibinfo {author} {\bibfnamefont {C.~E.}\ \bibnamefont
  {Granade}}, \bibinfo {author} {\bibfnamefont {C.}~\bibnamefont {Ferrie}},
  \bibinfo {author} {\bibfnamefont {N.}~\bibnamefont {Wiebe}},\ and\ \bibinfo
  {author} {\bibfnamefont {D.~G.}\ \bibnamefont {Cory}},\ }\bibfield  {title}
  {\bibinfo {title} {Robust online hamiltonian learning},\ }\href@noop {}
  {\bibfield  {journal} {\bibinfo  {journal} {New Journal of Physics}\ }\textbf
  {\bibinfo {volume} {14}},\ \bibinfo {pages} {103013} (\bibinfo {year}
  {2012})}\BibitemShut {NoStop}%
\bibitem [{\citenamefont {Chiribella}\ \emph
  {et~al.}(2008{\natexlab{a}})\citenamefont {Chiribella}, \citenamefont
  {D’Ariano},\ and\ \citenamefont {Perinotti}}]{chiribella2008quantum}%
  \BibitemOpen
  \bibfield  {author} {\bibinfo {author} {\bibfnamefont {G.}~\bibnamefont
  {Chiribella}}, \bibinfo {author} {\bibfnamefont {G.~M.}\ \bibnamefont
  {D’Ariano}},\ and\ \bibinfo {author} {\bibfnamefont {P.}~\bibnamefont
  {Perinotti}},\ }\bibfield  {title} {\bibinfo {title} {Quantum circuit
  architecture},\ }\href@noop {} {\bibfield  {journal} {\bibinfo  {journal}
  {Physical Review Letters}\ }\textbf {\bibinfo {volume} {101}},\ \bibinfo
  {pages} {060401} (\bibinfo {year} {2008}{\natexlab{a}})}\BibitemShut
  {NoStop}%
\bibitem [{\citenamefont {Chiribella}\ \emph
  {et~al.}(2008{\natexlab{b}})\citenamefont {Chiribella}, \citenamefont
  {D'Ariano},\ and\ \citenamefont {Perinotti}}]{chiribella2008transforming}%
  \BibitemOpen
  \bibfield  {author} {\bibinfo {author} {\bibfnamefont {G.}~\bibnamefont
  {Chiribella}}, \bibinfo {author} {\bibfnamefont {G.~M.}\ \bibnamefont
  {D'Ariano}},\ and\ \bibinfo {author} {\bibfnamefont {P.}~\bibnamefont
  {Perinotti}},\ }\bibfield  {title} {\bibinfo {title} {Transforming quantum
  operations: Quantum supermaps},\ }\href@noop {} {\bibfield  {journal}
  {\bibinfo  {journal} {Europhysics Letters}\ }\textbf {\bibinfo {volume}
  {83}},\ \bibinfo {pages} {30004} (\bibinfo {year}
  {2008}{\natexlab{b}})}\BibitemShut {NoStop}%
\bibitem [{\citenamefont {Chiribella}\ \emph {et~al.}(2009)\citenamefont
  {Chiribella}, \citenamefont {D’Ariano},\ and\ \citenamefont
  {Perinotti}}]{chiribella2009theoretical}%
  \BibitemOpen
  \bibfield  {author} {\bibinfo {author} {\bibfnamefont {G.}~\bibnamefont
  {Chiribella}}, \bibinfo {author} {\bibfnamefont {G.~M.}\ \bibnamefont
  {D’Ariano}},\ and\ \bibinfo {author} {\bibfnamefont {P.}~\bibnamefont
  {Perinotti}},\ }\bibfield  {title} {\bibinfo {title} {Theoretical framework
  for quantum networks},\ }\href@noop {} {\bibfield  {journal} {\bibinfo
  {journal} {Physical Review A}\ }\textbf {\bibinfo {volume} {80}},\ \bibinfo
  {pages} {022339} (\bibinfo {year} {2009})}\BibitemShut {NoStop}%
\bibitem [{\citenamefont {Bisio}\ and\ \citenamefont
  {Perinotti}(2019)}]{bisio2019theoretical}%
  \BibitemOpen
  \bibfield  {author} {\bibinfo {author} {\bibfnamefont {A.}~\bibnamefont
  {Bisio}}\ and\ \bibinfo {author} {\bibfnamefont {P.}~\bibnamefont
  {Perinotti}},\ }\bibfield  {title} {\bibinfo {title} {Theoretical framework
  for higher-order quantum theory},\ }\href@noop {} {\bibfield  {journal}
  {\bibinfo  {journal} {Proceedings of the Royal Society A}\ }\textbf {\bibinfo
  {volume} {475}},\ \bibinfo {pages} {20180706} (\bibinfo {year}
  {2019})}\BibitemShut {NoStop}%
\bibitem [{\citenamefont {Chitambar}\ and\ \citenamefont
  {Gour}(2019)}]{chitambar2019quantum}%
  \BibitemOpen
  \bibfield  {author} {\bibinfo {author} {\bibfnamefont {E.}~\bibnamefont
  {Chitambar}}\ and\ \bibinfo {author} {\bibfnamefont {G.}~\bibnamefont
  {Gour}},\ }\bibfield  {title} {\bibinfo {title} {Quantum resource theories},\
  }\href@noop {} {\bibfield  {journal} {\bibinfo  {journal} {Reviews of Modern
  Physics}\ }\textbf {\bibinfo {volume} {91}},\ \bibinfo {pages} {025001}
  (\bibinfo {year} {2019})}\BibitemShut {NoStop}%
\bibitem [{\citenamefont {Oreshkov}\ \emph {et~al.}(2012)\citenamefont
  {Oreshkov}, \citenamefont {Costa},\ and\ \citenamefont
  {Brukner}}]{oreshkov2012quantum}%
  \BibitemOpen
  \bibfield  {author} {\bibinfo {author} {\bibfnamefont {O.}~\bibnamefont
  {Oreshkov}}, \bibinfo {author} {\bibfnamefont {F.}~\bibnamefont {Costa}},\
  and\ \bibinfo {author} {\bibfnamefont {{\v{C}}.}~\bibnamefont {Brukner}},\
  }\bibfield  {title} {\bibinfo {title} {Quantum correlations with no causal
  order},\ }\href@noop {} {\bibfield  {journal} {\bibinfo  {journal} {Nature
  Communications}\ }\textbf {\bibinfo {volume} {3}},\ \bibinfo {pages} {1092}
  (\bibinfo {year} {2012})}\BibitemShut {NoStop}%
\bibitem [{\citenamefont {Miyazaki}\ \emph {et~al.}(2019)\citenamefont
  {Miyazaki}, \citenamefont {Soeda},\ and\ \citenamefont
  {Murao}}]{miyazaki2019complex}%
  \BibitemOpen
  \bibfield  {author} {\bibinfo {author} {\bibfnamefont {J.}~\bibnamefont
  {Miyazaki}}, \bibinfo {author} {\bibfnamefont {A.}~\bibnamefont {Soeda}},\
  and\ \bibinfo {author} {\bibfnamefont {M.}~\bibnamefont {Murao}},\ }\bibfield
   {title} {\bibinfo {title} {Complex conjugation supermap of unitary quantum
  maps and its universal implementation protocol},\ }\href@noop {} {\bibfield
  {journal} {\bibinfo  {journal} {Physical Review Research}\ }\textbf {\bibinfo
  {volume} {1}},\ \bibinfo {pages} {013007} (\bibinfo {year}
  {2019})}\BibitemShut {NoStop}%
\bibitem [{\citenamefont {Dong}\ \emph {et~al.}(2019)\citenamefont {Dong},
  \citenamefont {Nakayama}, \citenamefont {Soeda},\ and\ \citenamefont
  {Murao}}]{dong2019controlled}%
  \BibitemOpen
  \bibfield  {author} {\bibinfo {author} {\bibfnamefont {Q.}~\bibnamefont
  {Dong}}, \bibinfo {author} {\bibfnamefont {S.}~\bibnamefont {Nakayama}},
  \bibinfo {author} {\bibfnamefont {A.}~\bibnamefont {Soeda}},\ and\ \bibinfo
  {author} {\bibfnamefont {M.}~\bibnamefont {Murao}},\ }\bibfield  {title}
  {\bibinfo {title} {Controlled quantum operations and combs, and their
  applications to universal controllization of divisible unitary operations},\
  }\href@noop {} {\bibfield  {journal} {\bibinfo  {journal} {arXiv preprint
  arXiv:1911.01645}\ } (\bibinfo {year} {2019})}\BibitemShut {NoStop}%
\bibitem [{\citenamefont {Quintino}\ \emph
  {et~al.}(2019{\natexlab{a}})\citenamefont {Quintino}, \citenamefont {Dong},
  \citenamefont {Shimbo}, \citenamefont {Soeda},\ and\ \citenamefont
  {Murao}}]{quintino2019probabilistic}%
  \BibitemOpen
  \bibfield  {author} {\bibinfo {author} {\bibfnamefont {M.~T.}\ \bibnamefont
  {Quintino}}, \bibinfo {author} {\bibfnamefont {Q.}~\bibnamefont {Dong}},
  \bibinfo {author} {\bibfnamefont {A.}~\bibnamefont {Shimbo}}, \bibinfo
  {author} {\bibfnamefont {A.}~\bibnamefont {Soeda}},\ and\ \bibinfo {author}
  {\bibfnamefont {M.}~\bibnamefont {Murao}},\ }\bibfield  {title} {\bibinfo
  {title} {Probabilistic exact universal quantum circuits for transforming
  unitary operations},\ }\href@noop {} {\bibfield  {journal} {\bibinfo
  {journal} {Physical Review A}\ }\textbf {\bibinfo {volume} {100}},\ \bibinfo
  {pages} {062339} (\bibinfo {year} {2019}{\natexlab{a}})}\BibitemShut
  {NoStop}%
\bibitem [{\citenamefont {Quintino}\ \emph
  {et~al.}(2019{\natexlab{b}})\citenamefont {Quintino}, \citenamefont {Dong},
  \citenamefont {Shimbo}, \citenamefont {Soeda},\ and\ \citenamefont
  {Murao}}]{quintino2019reversing}%
  \BibitemOpen
  \bibfield  {author} {\bibinfo {author} {\bibfnamefont {M.~T.}\ \bibnamefont
  {Quintino}}, \bibinfo {author} {\bibfnamefont {Q.}~\bibnamefont {Dong}},
  \bibinfo {author} {\bibfnamefont {A.}~\bibnamefont {Shimbo}}, \bibinfo
  {author} {\bibfnamefont {A.}~\bibnamefont {Soeda}},\ and\ \bibinfo {author}
  {\bibfnamefont {M.}~\bibnamefont {Murao}},\ }\bibfield  {title} {\bibinfo
  {title} {Reversing unknown quantum transformations: Universal quantum circuit
  for inverting general unitary operations},\ }\href@noop {} {\bibfield
  {journal} {\bibinfo  {journal} {Physical Review Letters}\ }\textbf {\bibinfo
  {volume} {123}},\ \bibinfo {pages} {210502} (\bibinfo {year}
  {2019}{\natexlab{b}})}\BibitemShut {NoStop}%
\bibitem [{\citenamefont {Yoshida}\ \emph {et~al.}(2022)\citenamefont
  {Yoshida}, \citenamefont {Soeda},\ and\ \citenamefont
  {Murao}}]{yoshida2022reversing}%
  \BibitemOpen
  \bibfield  {author} {\bibinfo {author} {\bibfnamefont {S.}~\bibnamefont
  {Yoshida}}, \bibinfo {author} {\bibfnamefont {A.}~\bibnamefont {Soeda}},\
  and\ \bibinfo {author} {\bibfnamefont {M.}~\bibnamefont {Murao}},\ }\bibfield
   {title} {\bibinfo {title} {Reversing unknown qubit-unitary operation,
  deterministically and exactly},\ }\href@noop {} {\bibfield  {journal}
  {\bibinfo  {journal} {arXiv preprint arXiv:2209.02907}\ } (\bibinfo {year}
  {2022})}\BibitemShut {NoStop}%
\bibitem [{\citenamefont {Chiribella}\ and\ \citenamefont
  {Kristj{\'a}nsson}(2019)}]{chiribella2019quantum}%
  \BibitemOpen
  \bibfield  {author} {\bibinfo {author} {\bibfnamefont {G.}~\bibnamefont
  {Chiribella}}\ and\ \bibinfo {author} {\bibfnamefont {H.}~\bibnamefont
  {Kristj{\'a}nsson}},\ }\bibfield  {title} {\bibinfo {title} {Quantum
  {S}hannon theory with superpositions of trajectories},\ }\href@noop {}
  {\bibfield  {journal} {\bibinfo  {journal} {Proceedings of the Royal Society
  A}\ }\textbf {\bibinfo {volume} {475}},\ \bibinfo {pages} {20180903}
  (\bibinfo {year} {2019})}\BibitemShut {NoStop}%
\bibitem [{\citenamefont {Chiribella}\ \emph {et~al.}(2013)\citenamefont
  {Chiribella}, \citenamefont {D’Ariano}, \citenamefont {Perinotti},\ and\
  \citenamefont {Valiron}}]{chiribella2013quantum}%
  \BibitemOpen
  \bibfield  {author} {\bibinfo {author} {\bibfnamefont {G.}~\bibnamefont
  {Chiribella}}, \bibinfo {author} {\bibfnamefont {G.~M.}\ \bibnamefont
  {D’Ariano}}, \bibinfo {author} {\bibfnamefont {P.}~\bibnamefont
  {Perinotti}},\ and\ \bibinfo {author} {\bibfnamefont {B.}~\bibnamefont
  {Valiron}},\ }\bibfield  {title} {\bibinfo {title} {Quantum computations
  without definite causal structure},\ }\href@noop {} {\bibfield  {journal}
  {\bibinfo  {journal} {Physical Review A}\ }\textbf {\bibinfo {volume} {88}},\
  \bibinfo {pages} {022318} (\bibinfo {year} {2013})}\BibitemShut {NoStop}%
\bibitem [{\citenamefont {Pollock}\ \emph {et~al.}(2018)\citenamefont
  {Pollock}, \citenamefont {Rodr{\'\i}guez-Rosario}, \citenamefont
  {Frauenheim}, \citenamefont {Paternostro},\ and\ \citenamefont
  {Modi}}]{pollock2018non}%
  \BibitemOpen
  \bibfield  {author} {\bibinfo {author} {\bibfnamefont {F.~A.}\ \bibnamefont
  {Pollock}}, \bibinfo {author} {\bibfnamefont {C.}~\bibnamefont
  {Rodr{\'\i}guez-Rosario}}, \bibinfo {author} {\bibfnamefont {T.}~\bibnamefont
  {Frauenheim}}, \bibinfo {author} {\bibfnamefont {M.}~\bibnamefont
  {Paternostro}},\ and\ \bibinfo {author} {\bibfnamefont {K.}~\bibnamefont
  {Modi}},\ }\bibfield  {title} {\bibinfo {title} {Non-markovian quantum
  processes: Complete framework and efficient characterization},\ }\href@noop
  {} {\bibfield  {journal} {\bibinfo  {journal} {Physical Review A}\ }\textbf
  {\bibinfo {volume} {97}},\ \bibinfo {pages} {012127} (\bibinfo {year}
  {2018})}\BibitemShut {NoStop}%
\bibitem [{\citenamefont {Bai}\ \emph {et~al.}(2020)\citenamefont {Bai},
  \citenamefont {Wu}, \citenamefont {Zhu}, \citenamefont {Hayashi},\ and\
  \citenamefont {Chiribella}}]{bai2020efficient}%
  \BibitemOpen
  \bibfield  {author} {\bibinfo {author} {\bibfnamefont {G.}~\bibnamefont
  {Bai}}, \bibinfo {author} {\bibfnamefont {Y.-D.}\ \bibnamefont {Wu}},
  \bibinfo {author} {\bibfnamefont {Y.}~\bibnamefont {Zhu}}, \bibinfo {author}
  {\bibfnamefont {M.}~\bibnamefont {Hayashi}},\ and\ \bibinfo {author}
  {\bibfnamefont {G.}~\bibnamefont {Chiribella}},\ }\bibfield  {title}
  {\bibinfo {title} {Efficient algorithms for causal order discovery in quantum
  networks},\ }\href@noop {} {\bibfield  {journal} {\bibinfo  {journal} {arXiv
  preprint arXiv:2012.01731}\ } (\bibinfo {year} {2020})}\BibitemShut {NoStop}%
\bibitem [{\citenamefont {Gavorov{\'a}}\ \emph {et~al.}(2020)\citenamefont
  {Gavorov{\'a}}, \citenamefont {Seidel},\ and\ \citenamefont
  {Touati}}]{gavorova2020topological}%
  \BibitemOpen
  \bibfield  {author} {\bibinfo {author} {\bibfnamefont {Z.}~\bibnamefont
  {Gavorov{\'a}}}, \bibinfo {author} {\bibfnamefont {M.}~\bibnamefont
  {Seidel}},\ and\ \bibinfo {author} {\bibfnamefont {Y.}~\bibnamefont
  {Touati}},\ }\bibfield  {title} {\bibinfo {title} {Topological obstructions
  to implementing controlled unknown unitaries},\ }\href@noop {} {\bibfield
  {journal} {\bibinfo  {journal} {arXiv preprint arXiv:2011.10031}\ } (\bibinfo
  {year} {2020})}\BibitemShut {NoStop}%
\bibitem [{\citenamefont {Ara{\'u}jo}\ \emph {et~al.}(2014)\citenamefont
  {Ara{\'u}jo}, \citenamefont {Feix}, \citenamefont {Costa},\ and\
  \citenamefont {Brukner}}]{araujo2014quantum}%
  \BibitemOpen
  \bibfield  {author} {\bibinfo {author} {\bibfnamefont {M.}~\bibnamefont
  {Ara{\'u}jo}}, \bibinfo {author} {\bibfnamefont {A.}~\bibnamefont {Feix}},
  \bibinfo {author} {\bibfnamefont {F.}~\bibnamefont {Costa}},\ and\ \bibinfo
  {author} {\bibfnamefont {{\v{C}}.}~\bibnamefont {Brukner}},\ }\bibfield
  {title} {\bibinfo {title} {Quantum circuits cannot control unknown
  operations},\ }\href@noop {} {\bibfield  {journal} {\bibinfo  {journal} {New
  Journal of Physics}\ }\textbf {\bibinfo {volume} {16}},\ \bibinfo {pages}
  {093026} (\bibinfo {year} {2014})}\BibitemShut {NoStop}%
\bibitem [{\citenamefont {Soeda}(2013)}]{soeda2013limitations}%
  \BibitemOpen
  \bibfield  {author} {\bibinfo {author} {\bibfnamefont {A.}~\bibnamefont
  {Soeda}},\ }\href@noop {} {\emph {\bibinfo {title} {Limitations on quantum
  subroutine designing due to the linear structure of quantum operators}}}\
  (\bibinfo  {publisher} {Talk at the International Conference on Quantum
  Information and Technology (IC- QIT)},\ \bibinfo {year} {2013})\BibitemShut
  {NoStop}%
\bibitem [{\citenamefont {Friis}\ \emph {et~al.}(2014)\citenamefont {Friis},
  \citenamefont {Dunjko}, \citenamefont {D{\"u}r},\ and\ \citenamefont
  {Briegel}}]{friis2014implementing}%
  \BibitemOpen
  \bibfield  {author} {\bibinfo {author} {\bibfnamefont {N.}~\bibnamefont
  {Friis}}, \bibinfo {author} {\bibfnamefont {V.}~\bibnamefont {Dunjko}},
  \bibinfo {author} {\bibfnamefont {W.}~\bibnamefont {D{\"u}r}},\ and\ \bibinfo
  {author} {\bibfnamefont {H.~J.}\ \bibnamefont {Briegel}},\ }\bibfield
  {title} {\bibinfo {title} {Implementing quantum control for unknown
  subroutines},\ }\href@noop {} {\bibfield  {journal} {\bibinfo  {journal}
  {Physical Review A}\ }\textbf {\bibinfo {volume} {89}},\ \bibinfo {pages}
  {030303} (\bibinfo {year} {2014})}\BibitemShut {NoStop}%
\bibitem [{\citenamefont {Nakayama}\ \emph {et~al.}(2015)\citenamefont
  {Nakayama}, \citenamefont {Soeda},\ and\ \citenamefont
  {Murao}}]{nakayama2015quantum}%
  \BibitemOpen
  \bibfield  {author} {\bibinfo {author} {\bibfnamefont {S.}~\bibnamefont
  {Nakayama}}, \bibinfo {author} {\bibfnamefont {A.}~\bibnamefont {Soeda}},\
  and\ \bibinfo {author} {\bibfnamefont {M.}~\bibnamefont {Murao}},\ }\bibfield
   {title} {\bibinfo {title} {Quantum algorithm for universal implementation of
  the projective measurement of energy},\ }\href@noop {} {\bibfield  {journal}
  {\bibinfo  {journal} {Physical Review Letters}\ }\textbf {\bibinfo {volume}
  {114}},\ \bibinfo {pages} {190501} (\bibinfo {year} {2015})}\BibitemShut
  {NoStop}%
\bibitem [{\citenamefont {Chow}\ \emph {et~al.}(2012)\citenamefont {Chow},
  \citenamefont {Gambetta}, \citenamefont {Corcoles}, \citenamefont {Merkel},
  \citenamefont {Smolin}, \citenamefont {Rigetti}, \citenamefont {Poletto},
  \citenamefont {Keefe}, \citenamefont {Rothwell}, \citenamefont {Rozen} \emph
  {et~al.}}]{chow2012universal}%
  \BibitemOpen
  \bibfield  {author} {\bibinfo {author} {\bibfnamefont {J.~M.}\ \bibnamefont
  {Chow}}, \bibinfo {author} {\bibfnamefont {J.~M.}\ \bibnamefont {Gambetta}},
  \bibinfo {author} {\bibfnamefont {A.~D.}\ \bibnamefont {Corcoles}}, \bibinfo
  {author} {\bibfnamefont {S.~T.}\ \bibnamefont {Merkel}}, \bibinfo {author}
  {\bibfnamefont {J.~A.}\ \bibnamefont {Smolin}}, \bibinfo {author}
  {\bibfnamefont {C.}~\bibnamefont {Rigetti}}, \bibinfo {author} {\bibfnamefont
  {S.}~\bibnamefont {Poletto}}, \bibinfo {author} {\bibfnamefont {G.~A.}\
  \bibnamefont {Keefe}}, \bibinfo {author} {\bibfnamefont {M.~B.}\ \bibnamefont
  {Rothwell}}, \bibinfo {author} {\bibfnamefont {J.~R.}\ \bibnamefont {Rozen}},
  \emph {et~al.},\ }\bibfield  {title} {\bibinfo {title} {Universal quantum
  gate set approaching fault-tolerant thresholds with superconducting qubits},\
  }\href@noop {} {\bibfield  {journal} {\bibinfo  {journal} {Physical review
  letters}\ }\textbf {\bibinfo {volume} {109}},\ \bibinfo {pages} {060501}
  (\bibinfo {year} {2012})}\BibitemShut {NoStop}%
\bibitem [{\citenamefont {Akibue}\ \emph {et~al.}(2023)\citenamefont {Akibue},
  \citenamefont {Kato},\ and\ \citenamefont {Tani}}]{Akibue}%
  \BibitemOpen
  \bibfield  {author} {\bibinfo {author} {\bibfnamefont {S.}~\bibnamefont
  {Akibue}}, \bibinfo {author} {\bibfnamefont {G.}~\bibnamefont {Kato}},\ and\
  \bibinfo {author} {\bibfnamefont {S.}~\bibnamefont {Tani}},\ }\bibfield
  {title} {\bibinfo {title} {Probabilistic state synthesis based on optimal
  convex approximation},\ }\href@noop {} {\bibfield  {journal} {\bibinfo
  {journal} {arXiv preprint arXiv:2303.10860}\ } (\bibinfo {year}
  {2023})}\BibitemShut {NoStop}%
\bibitem [{\citenamefont {Huang}\ \emph {et~al.}(2023)\citenamefont {Huang},
  \citenamefont {Tong}, \citenamefont {Fang},\ and\ \citenamefont
  {Su}}]{huang2023learning}%
  \BibitemOpen
  \bibfield  {author} {\bibinfo {author} {\bibfnamefont {H.-Y.}\ \bibnamefont
  {Huang}}, \bibinfo {author} {\bibfnamefont {Y.}~\bibnamefont {Tong}},
  \bibinfo {author} {\bibfnamefont {D.}~\bibnamefont {Fang}},\ and\ \bibinfo
  {author} {\bibfnamefont {Y.}~\bibnamefont {Su}},\ }\bibfield  {title}
  {\bibinfo {title} {Learning many-body hamiltonians with heisenberg-limited
  scaling},\ }\href@noop {} {\bibfield  {journal} {\bibinfo  {journal}
  {Physical Review Letters}\ }\textbf {\bibinfo {volume} {130}},\ \bibinfo
  {pages} {200403} (\bibinfo {year} {2023})}\BibitemShut {NoStop}%
\bibitem [{\citenamefont {Weinberg}(2005)}]{weinberg}%
  \BibitemOpen
  \bibfield  {author} {\bibinfo {author} {\bibfnamefont {S.}~\bibnamefont
  {Weinberg}},\ }\href@noop {} {\emph {\bibinfo {title} {{The Quantum Theory of
  Fields, Volume 1: Foundations}}}}\ (\bibinfo  {publisher} {Cambridge
  University Press},\ \bibinfo {year} {2005})\BibitemShut {NoStop}%
\bibitem [{\citenamefont {Sakurai}(1994)}]{Sakurai}%
  \BibitemOpen
  \bibfield  {author} {\bibinfo {author} {\bibfnamefont {J.~J.}\ \bibnamefont
  {Sakurai}},\ }\href {https://cds.cern.ch/record/1167961} {\emph {\bibinfo
  {title} {{Modern quantum mechanics; rev. ed.}}}}\ (\bibinfo  {publisher}
  {Addison-Wesley},\ \bibinfo {address} {Reading, MA},\ \bibinfo {year}
  {1994})\BibitemShut {NoStop}%
\bibitem [{\citenamefont {de~Burgh}\ and\ \citenamefont
  {Bartlett}(2005)}]{de2005quantum}%
  \BibitemOpen
  \bibfield  {author} {\bibinfo {author} {\bibfnamefont {M.}~\bibnamefont
  {de~Burgh}}\ and\ \bibinfo {author} {\bibfnamefont {S.~D.}\ \bibnamefont
  {Bartlett}},\ }\bibfield  {title} {\bibinfo {title} {Quantum methods for
  clock synchronization: Beating the standard quantum limit without
  entanglement},\ }\href@noop {} {\bibfield  {journal} {\bibinfo  {journal}
  {Physical Review A}\ }\textbf {\bibinfo {volume} {72}},\ \bibinfo {pages}
  {042301} (\bibinfo {year} {2005})}\BibitemShut {NoStop}%
\bibitem [{\citenamefont {Wiebe}\ \emph {et~al.}(2014)\citenamefont {Wiebe},
  \citenamefont {Granade}, \citenamefont {Ferrie},\ and\ \citenamefont
  {Cory}}]{wiebe2014quantum}%
  \BibitemOpen
  \bibfield  {author} {\bibinfo {author} {\bibfnamefont {N.}~\bibnamefont
  {Wiebe}}, \bibinfo {author} {\bibfnamefont {C.}~\bibnamefont {Granade}},
  \bibinfo {author} {\bibfnamefont {C.}~\bibnamefont {Ferrie}},\ and\ \bibinfo
  {author} {\bibfnamefont {D.}~\bibnamefont {Cory}},\ }\bibfield  {title}
  {\bibinfo {title} {Quantum hamiltonian learning using imperfect quantum
  resources},\ }\href@noop {} {\bibfield  {journal} {\bibinfo  {journal}
  {Physical Review A}\ }\textbf {\bibinfo {volume} {89}},\ \bibinfo {pages}
  {042314} (\bibinfo {year} {2014})}\BibitemShut {NoStop}%
\bibitem [{\citenamefont {Boulant}\ \emph {et~al.}(2003)\citenamefont
  {Boulant}, \citenamefont {Havel}, \citenamefont {Pravia},\ and\ \citenamefont
  {Cory}}]{boulant2003robust}%
  \BibitemOpen
  \bibfield  {author} {\bibinfo {author} {\bibfnamefont {N.}~\bibnamefont
  {Boulant}}, \bibinfo {author} {\bibfnamefont {T.~F.}\ \bibnamefont {Havel}},
  \bibinfo {author} {\bibfnamefont {M.~A.}\ \bibnamefont {Pravia}},\ and\
  \bibinfo {author} {\bibfnamefont {D.~G.}\ \bibnamefont {Cory}},\ }\bibfield
  {title} {\bibinfo {title} {Robust method for estimating the lindblad
  operators of a dissipative quantum process from measurements of the density
  operator at multiple time points},\ }\href@noop {} {\bibfield  {journal}
  {\bibinfo  {journal} {Physical Review A}\ }\textbf {\bibinfo {volume} {67}},\
  \bibinfo {pages} {042322} (\bibinfo {year} {2003})}\BibitemShut {NoStop}%
\bibitem [{\citenamefont {Kimmel}\ \emph {et~al.}(2015)\citenamefont {Kimmel},
  \citenamefont {Low},\ and\ \citenamefont {Yoder}}]{kimmel2015robust}%
  \BibitemOpen
  \bibfield  {author} {\bibinfo {author} {\bibfnamefont {S.}~\bibnamefont
  {Kimmel}}, \bibinfo {author} {\bibfnamefont {G.~H.}\ \bibnamefont {Low}},\
  and\ \bibinfo {author} {\bibfnamefont {T.~J.}\ \bibnamefont {Yoder}},\
  }\bibfield  {title} {\bibinfo {title} {Robust calibration of a universal
  single-qubit gate set via robust phase estimation},\ }\href@noop {}
  {\bibfield  {journal} {\bibinfo  {journal} {Physical Review A}\ }\textbf
  {\bibinfo {volume} {92}},\ \bibinfo {pages} {062315} (\bibinfo {year}
  {2015})}\BibitemShut {NoStop}%
\bibitem [{\citenamefont {Watrous}(2018)}]{watrous2018theory}%
  \BibitemOpen
  \bibfield  {author} {\bibinfo {author} {\bibfnamefont {J.}~\bibnamefont
  {Watrous}},\ }\href@noop {} {\emph {\bibinfo {title} {The theory of quantum
  information}}}\ (\bibinfo  {publisher} {Cambridge University Press},\
  \bibinfo {year} {2018})\BibitemShut {NoStop}%
\bibitem [{\citenamefont {Fuchs}\ and\ \citenamefont {Van
  De~Graaf}(1999)}]{fuchs1999cryptographic}%
  \BibitemOpen
  \bibfield  {author} {\bibinfo {author} {\bibfnamefont {C.~A.}\ \bibnamefont
  {Fuchs}}\ and\ \bibinfo {author} {\bibfnamefont {J.}~\bibnamefont {Van
  De~Graaf}},\ }\bibfield  {title} {\bibinfo {title} {Cryptographic
  distinguishability measures for quantum-mechanical states},\ }\href@noop {}
  {\bibfield  {journal} {\bibinfo  {journal} {IEEE Transactions on Information
  Theory}\ }\textbf {\bibinfo {volume} {45}},\ \bibinfo {pages} {1216}
  (\bibinfo {year} {1999})}\BibitemShut {NoStop}%
\bibitem [{\citenamefont {Gily{\'e}n}\ \emph {et~al.}(2019)\citenamefont
  {Gily{\'e}n}, \citenamefont {Su}, \citenamefont {Low},\ and\ \citenamefont
  {Wiebe}}]{gilyen2019quantum}%
  \BibitemOpen
  \bibfield  {author} {\bibinfo {author} {\bibfnamefont {A.}~\bibnamefont
  {Gily{\'e}n}}, \bibinfo {author} {\bibfnamefont {Y.}~\bibnamefont {Su}},
  \bibinfo {author} {\bibfnamefont {G.~H.}\ \bibnamefont {Low}},\ and\ \bibinfo
  {author} {\bibfnamefont {N.}~\bibnamefont {Wiebe}},\ }\bibfield  {title}
  {\bibinfo {title} {Quantum singular value transformation and beyond:
  exponential improvements for quantum matrix arithmetics},\ }in\ \href@noop {}
  {\emph {\bibinfo {booktitle} {Proceedings of the 51st Annual ACM SIGACT
  Symposium on Theory of Computing}}}\ (\bibinfo {year} {2019})\ pp.\ \bibinfo
  {pages} {193--204}\BibitemShut {NoStop}%
\bibitem [{\citenamefont {Martyn}\ \emph {et~al.}(2021)\citenamefont {Martyn},
  \citenamefont {Rossi}, \citenamefont {Tan},\ and\ \citenamefont
  {Chuang}}]{martyn2021grand}%
  \BibitemOpen
  \bibfield  {author} {\bibinfo {author} {\bibfnamefont {J.~M.}\ \bibnamefont
  {Martyn}}, \bibinfo {author} {\bibfnamefont {Z.~M.}\ \bibnamefont {Rossi}},
  \bibinfo {author} {\bibfnamefont {A.~K.}\ \bibnamefont {Tan}},\ and\ \bibinfo
  {author} {\bibfnamefont {I.~L.}\ \bibnamefont {Chuang}},\ }\bibfield  {title}
  {\bibinfo {title} {Grand unification of quantum algorithms},\ }\href@noop {}
  {\bibfield  {journal} {\bibinfo  {journal} {PRX Quantum}\ }\textbf {\bibinfo
  {volume} {2}},\ \bibinfo {pages} {040203} (\bibinfo {year}
  {2021})}\BibitemShut {NoStop}%
\bibitem [{\citenamefont {Haah}\ \emph {et~al.}(2023)\citenamefont {Haah},
  \citenamefont {Kothari}, \citenamefont {O'Donnell},\ and\ \citenamefont
  {Tang}}]{haah2023query}%
  \BibitemOpen
  \bibfield  {author} {\bibinfo {author} {\bibfnamefont {J.}~\bibnamefont
  {Haah}}, \bibinfo {author} {\bibfnamefont {R.}~\bibnamefont {Kothari}},
  \bibinfo {author} {\bibfnamefont {R.}~\bibnamefont {O'Donnell}},\ and\
  \bibinfo {author} {\bibfnamefont {E.}~\bibnamefont {Tang}},\ }\bibfield
  {title} {\bibinfo {title} {Query-optimal estimation of unitary channels in
  diamond distance},\ }\href@noop {} {\bibfield  {journal} {\bibinfo  {journal}
  {arXiv preprint arXiv:2302.14066}\ } (\bibinfo {year} {2023})}\BibitemShut
  {NoStop}%
\end{thebibliography}
\end{document}